\documentclass[10pt,conference]{IEEEtran}

\usepackage{etoolbox}
\makeatletter
\def\ps@headings{%
	\def\@oddhead{\mbox{}\scriptsize\rightmark \hfil \thepage}%
	\def\@evenhead{\scriptsize\thepage \hfil \leftmark\mbox{}}%
	\def\@oddfoot{}%
	\def\@evenfoot{}}
\makeatother
\usepackage{verbatim}
\usepackage{xcolor}
\usepackage{amsfonts}
\usepackage{mathrsfs}
\usepackage{amsfonts}
\usepackage{amssymb}
\usepackage{graphicx}
\usepackage{epsfig}
\usepackage{epstopdf}
\usepackage{psfrag}
\usepackage{amsmath}
\usepackage{array}
\usepackage{cases}
\usepackage{subfigure}
\usepackage{cite,graphicx,amsmath,amssymb,color}
\usepackage{algorithmic}
\usepackage{algorithm}
\usepackage{multicol}

\usepackage{algorithm}
\usepackage{algorithmic}
\usepackage{stmaryrd}
\usepackage{multirow}
\usepackage{subfig}
\usepackage{graphicx,times,amsmath} 

\usepackage{url}
\usepackage{enumerate}
\usepackage{caption}
\IEEEoverridecommandlockouts

\newtheorem{theorem}{Theorem}

\newtheorem{Exam}{Example}

\newtheorem{problem}{Problem}

\IEEEoverridecommandlockouts
\begin{document}
\bibliographystyle{IEEEtran}

\title{Optimal Wireless Streaming of Multi-Quality 360 VR Video by Exploiting Natural, Relative Smoothness-enabled and Transcoding-enabled Multicast Opportunities}
\author{\IEEEauthorblockN{Kaixuan Long, \quad Ying Cui, \quad Chencheng Ye, \quad Zhi Liu}\thanks{K. Long, Y. Cui and C. Ye are with Shanghai Jiao Tong University, China. Z. Liu is with Shizuoka  University, Japan. This paper was presented in part at the IEEE GLOBECOM 2019~\cite{long2020optimal}.}}
\maketitle

\begin{abstract}
  In this paper, we would like to investigate optimal wireless streaming of a multi-quality tiled 360 virtual reality (VR) video from a server to multiple users. To this end, we propose to maximally exploit potential multicast opportunities by effectively utilizing characteristics of multi-quality tiled 360 VR videos and computation resources at the users’ side. In particular, we consider two requirements for quality variation in one field-of-view (FoV), i.e., the absolute smoothness requirement and the relative smoothness requirement, and two video playback modes, i.e., the direct-playback mode (without user transcoding) and transcode-playback mode (with user transcoding). Besides natural multicast opportunities, we introduce two new types of multicast opportunities, namely, relative smoothness-enabled multicast opportunities, which allow flexible tradeoff between viewing quality and communications resource consumption, and transcoding-enabled multicast opportunities, which allow flexible tradeoff between computation and communications resource consumptions. Then, we establish a novel mathematical model that reflects the impacts of natural, relative smoothness-enabled and transcoding-enabled multicast opportunities on the average transmission energy and transcoding energy. Based on this model, we optimize the transmission resource allocation, playback quality level selection and transmission quality level selection to minimize the energy consumption in the four cases with different requirements for quality variation and video playback modes. By comparing the optimal values in the four cases, we prove that the energy consumption reduces when more multicast opportunities can be utilized. Finally, numerical results show substantial gains of the proposed solutions over existing schemes, and demonstrate the importance of effective exploitation of the three types of multicast opportunities.
\end{abstract}



\section{Introduction}\label{section_1}

Virtual reality (VR) video is generated by capturing a scene of interest in every direction at the same time using omnidirectional cameras.
A captured video is stitched and warped onto a 3D sphere, and then projected onto a 2D map using projection methods such as equirectangular projection, cubemap projection and pyramid projection.
The resulting video is referred to as 360 VR video.
The most commonly used projection method is equirectangular projection which projects a 3D sphere onto a rectangle~\cite{zink2019scalable}.
A user wearing a VR headset or head mounted display (HMD) can freely watch the scene of interest in any viewing direction at any time, hence enjoying immersive viewing experience. VR has vast applications in entertainment, education, medicine, etc.
VR videos are becoming increasingly accessible, as several large video sharing websites and social media platforms have started providing VR services.
It is predicted that the VR market will reach 87.97 billion USD by 2025~\cite{111}.

Most existing VR headsets connect to personal computers (PCs) via cables, which greatly limits user mobility and experience.
Increasing efforts have been devoted to wireless transmission of 360 VR videos.
A 360 VR video is of a much larger size than a traditional video~\cite{song2019fast}.
Thus, transmitting an entire 360 VR video brings a heavy burden to wireless networks. In addition, at any moment a user watching a 360 VR video is interested in only one viewing direction. Thus, transmitting an entire 360 VR video is also unnecessary.
To improve transmission efficiency for 360 VR videos, tiling technique is widely adopted~\cite{gaddam2016tiling,maniotis2019tile,song2019fast}.
Specifically, a 360 VR video is divided into smaller rectangular segments of the same size, referred to as tiles.
Transmitting the set of tiles covering a user's field-of-view (FoV) can save communications resource, without degrading the user's quality of experience (QoE).
This relies on viewing direction (or FoV) prediction, as a VR user may change viewing directions from time to time.
Commonly adopted viewing direction prediction methods deal with the prediction of an individual user's viewing direction based on his historical head movement~\cite{Qian:2016:OVD:2980055.2980056,qian2018flare,bao2016shooting,fan2019optimizing}, or the prediction of popular viewing directions based on the head movement trajectories of multiple users~\cite{liu2017360,DBLP:journals/corr/abs-1803-08177}.
Typically, the angular rate of human's head rotation is  limited (below $100^{\circ}/s$)~\cite{Ju:2017:UWV:3097895.3097899}. Thus, viewing direction can be well predicted in a time interval of 0.1s-0.5s.
The prediction accuracy increases as the prediction time interval reduces.
To deal with possible prediction errors under tiling, two basic transmission schemes are commonly adopted. One is to transmit the tiles in the predicted FoV plus a safe margin at the desired quality~\cite{zhou2018clustile,DBLP:journals/corr/CheungLMT17,8428401,8478317,Xie:2017:IQV:3123266.3123291,He2018Joint,nguyen2019optimal,DBLP:journals/corr/abs-1901-02203}. Note that the size of the safe margin can adapt to the FoV prediction accuracy. The other is to transmit the tiles in the predicted FoV at the desired quality and the remaining tiles at low quality~\cite{7996611,ozcinar2017viewport,rossi2017navigation,xiao2018bas,liu2018jet,Ahmadi:2017:AMS:3126686.3126743}. The former one has higher spectrum efficiency than the latter one, and can provide tiles of uniform quality, when the prediction accuracy is not quite low. In contrast, the latter one can guarantee that the tiles in the actual FoV can always be delivered.
Obviously, such two basic transmission schemes can be combined to achieve complementary advantages. In particular, one can transmit the tiles in the predicted FoV plus a safe margin at the desired quality and the remaining tiles at low quality. When the size of the margin reduces to zero, the combined transmission scheme degenerates to the second basic transmission scheme.

In~\cite{zhou2018clustile,DBLP:journals/corr/CheungLMT17}, the authors consider streaming of single-quality 360 VR videos in single-user wireless networks.
The proposed solutions in \cite{zhou2018clustile,DBLP:journals/corr/CheungLMT17} may not yield efficient transmission for single-quality 360 VR videos in multi-user wireless networks, as optimal resource sharing among users with heterogeneous channel conditions is not considered.
Furthermore, when multiple users are watching one 360 VR video simultaneously, transmission efficiency can be improved by exploiting potential multicast opportunities.
In our previous works~\cite{8428401,8478317}, we consider optimal streaming of a single-quality tiled 360 VR video in a Time Division Multiple Access (TDMA) system and an Orthogonal Frequency Division Multiple Access (OFDMA) system, respectively, by exploiting natural multicast opportunities.
Specifically, we consider the optimal transmission resource allocation to minimize the average transmission energy for given video quality requirements of all users;
and we also consider the optimization of the encoding rate of each tile to maximize the received video quality for a given transmission energy budget for the serving node.

In~\cite{Xie:2017:IQV:3123266.3123291,He2018Joint,nguyen2019optimal,7996611,ozcinar2017viewport,rossi2017navigation,xiao2018bas,liu2018jet}, the authors consider streaming of multi-quality tiled 360 VR videos in singer-user wireless networks.
The main focus of~\cite{Xie:2017:IQV:3123266.3123291,He2018Joint,nguyen2019optimal,7996611,ozcinar2017viewport,rossi2017navigation,xiao2018bas,liu2018jet} is the quality level selection for each tile to be transmitted.
Specifically, the proposed schemes in \cite{7996611,ozcinar2017viewport} are heuristic, while those in \cite{Xie:2017:IQV:3123266.3123291,He2018Joint,nguyen2019optimal,rossi2017navigation,xiao2018bas,liu2018jet} are optimization based, with distortion, bandwidth, utility, etc. being the objective functions or constraint functions.
Similarly, the proposed solutions in~\cite{7996611,ozcinar2017viewport,Xie:2017:IQV:3123266.3123291,rossi2017navigation,xiao2018bas,He2018Joint,liu2018jet,nguyen2019optimal} may not result in efficient transmission design for multi-quality tiled 360 VR videos in multi-user wireless networks, especially for the case where multiple users are watching the same 360 VR video.
In~\cite{Ahmadi:2017:AMS:3126686.3126743,DBLP:journals/corr/abs-1901-02203}, wireless streaming of a multi-quality tiled 360 VR video to multiple users is considered, and quality variation for tiles in an FoV is allowed so that more multicast opportunities can be exploited for efficient transmission.
Specifically,~\cite{Ahmadi:2017:AMS:3126686.3126743} optimizes the quality level selection for each tile to be transmitted to maximize the total utility of all users under some communications resource constraints. The size of the optimization problem is unnecessarily large, as tiles are considered separately. In addition, without any constraints on quality variation, the obtained quality levels of adjacent tiles may vary significantly, leading to poor viewing experience.
In our previous work~\cite{DBLP:journals/corr/abs-1901-02203}, we study the optimal quality level selection to maximize the total utility of all users under communications resource constraints and quality smoothness constraints which limit the quality variation for any two adjacent tiles.
However, the quality smoothness constraints for adjacent tiles in~\cite{DBLP:journals/corr/abs-1901-02203} still cannot effectively control the level of quality variation in an FoV. Furthermore, the number of the smoothness constraints for adjacent tiles is huge, resulting in a substantial increase in the computational complexity for solving the optimization problem.
Besides,~\cite{Ahmadi:2017:AMS:3126686.3126743,DBLP:journals/corr/abs-1901-02203} neglect the fact that channel conditions of users change much faster than their FoVs, and hence the proposed solutions in~\cite{Ahmadi:2017:AMS:3126686.3126743,DBLP:journals/corr/abs-1901-02203} may not yield desired performance in practical systems.

In this paper, we would like to minimize the energy consumption for wireless streaming of a multi-quality tiled 360 VR video to multiple users. To this end, we propose to maximally exploit potential multicast opportunities by effectively utilizing characteristics of multi-quality tiled 360 VR videos and computation resources at the users’ side. In particular, we consider two requirements for quality variation in one FoV, i.e., the absolute smoothness requirement and the relative smoothness requirement, and two video playback modes, i.e., the direct-playback mode (without user transcoding) and transcode-playback mode (with user transcoding), and investigate potential multicast opportunities for optimal wireless streaming in the four cases with different  requirements for quality variation and video playback modes. The main contributions of this paper are summarized below.\footnote{This paper extends the results under the absolute smoothness requirement in the conference version~\cite{long2020optimal} to those under the relative smoothness requirement. Besides the two cases with the absolute smoothness requirement and the natural and the transcoding-enabled multicast opportunities investigated in~\cite{long2020optimal}, in this paper, we also study the two cases with the relative smoothness requirement and the relative smoothness-enabled multicast opportunities.}

\begin{itemize}
  \item We introduce an elegant notation system for partitioning all tiles into subsets, each for a particular group of users, and specifying the relation between a subset of tiles and their target user group. This new notation system is more tractable and intuitive than the one in our previous works~\cite{8428401,8478317,DBLP:journals/corr/abs-1901-02203}.
  \item We introduce two new types of multicast opportunities in transmission of the multi-quality tiled 360 VR video, namely, relative smoothness-enabled multicast opportunities, which allow flexible tradeoff between viewing quality and communications resource consumption,  and transcoding-enabled multicast opportunities, which allow flexible tradeoff between computation and communications resource consumptions. Furthermore, we establish a novel mathematical model that reflects the impacts of multicast opportunities on the average transmission energy and transcoding energy under controllable quality variation for tiles in an FoV, and thus greatly facilitates optimal exploitation of potential multicast opportunities for energy minimization. To the best of our knowledge, neither relative smoothness-enabled multicast opportunities nor transcoding-enabled multicast opportunities have been recognized to improve the efficiency of wireless streaming of a 360 VR video to multiple users.
  \item We minimize the average transmission energy in the two cases without user transcoding. In particular, under the absolute smoothness requirement, we optimize the transmission resource, leading to a non-convex problem. We develop an efficient algorithm to obtain an optimal solution using transformation techniques and convex optimization techniques. Under the relative smoothness requirement, we optimize the transmission resource and transmission quality level selection, resulting in a challenging mixed discrete-continuous optimization. We develop an algorithm to obtain a suboptimal solution by convex concave procedure. We also minimize the weighted sum of the average transmission energy and the transcoding energy in the two cases with user transcoding. Specifically, under the absolute smoothness requirement, we optimize the transmission resource allocation and transmission quality level selection, while under the relative smoothness requirement, we optimize the transmission resource allocation, playback quality level selection and transmission quality level selection. Both problems are challenging mixed discrete-continuous optimization problems, and we obtain their suboptimal solutions using  convex concave procedure. By comparing the optimal values in the four cases, we prove that the energy consumption reduces when more multicast opportunities can be utilized.
  \item Numerical results show substantial gains of the proposed solutions over existing schemes in all four cases, and demonstrate the importance of effective exploitation of the three types of multicast opportunities for energy efficient wireless streaming of a multi-quality tiled 360 VR video to multiple users.
\end{itemize}

\begin{figure*}[t]
\normalsize{
\centering
\subfigure{
\begin{minipage}{14.1cm}
\centering
\includegraphics[width=14.1cm]{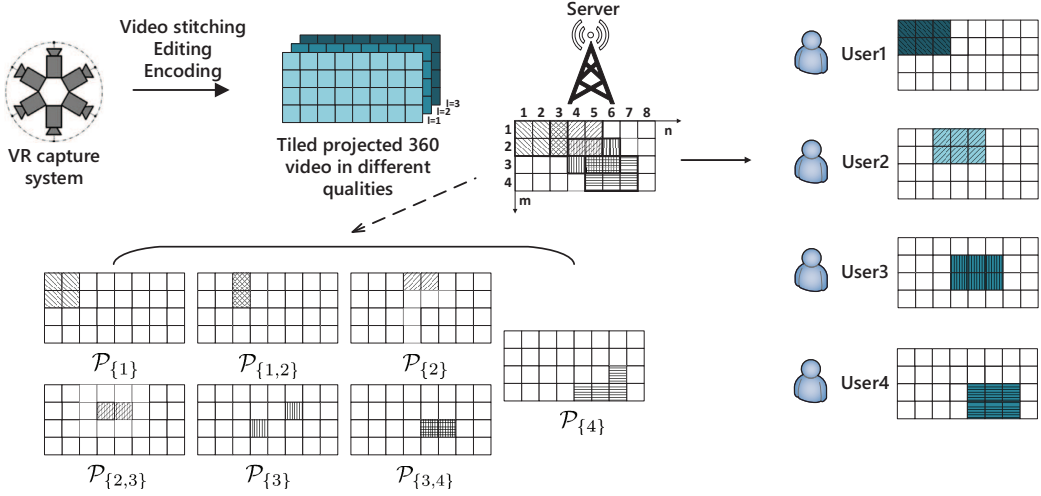}
\end{minipage}
}
\caption{  System model. $K$$=$$4,\,M$$=$$4,\,N$$=$$8,\,L$$=$$3,\,\mathcal{G}_{1}$$=$$\{(1,1),$$(2,1),$$(1,2),$$(2,2),$$(1,3),$$(2,3)\},\,\mathcal{G}_{2}$$=$$\{(1,3),$$(2,3),$$(1,4),$$(2,4),$$(1,5),$$(2,5)\},\,$
$\mathcal{G}_{3}$$=$$\{(2,4),$$(3,4),$$(2,5),$$(3,5),$$(2,6),$$(3,6)\},\,\mathcal{G}_{4}$$=\{(3,5),$$(4,5),$$(3,6),$$(4,6),$$(3,7),$$(4,7)\},$$r_{1}$$=$$3,\,r_{2}$$=$$1, \,r_{3}$$=$$2$ and $r_{4}$$=$$2$.}
\label{system0}
}
\end{figure*}

\section{System Model}\label{section_2}

As illustrated in Fig.~\ref{system0}, we consider wireless streaming of a multi-quality tiled 360 VR video from a single-antenna server (e.g., base station or access point) to $K$ $(\geq1)$ single-antenna users, each wearing a VR headset, in a TDMA system.\footnote{TDMA is more analytically tractable and has applications in WiFi systems. The proposed multicast transmission schemes and optimization frameworks for TDMA systems can be extended to OFDMA systems, multi-user MIMO systems, etc.} Let $\mathcal{K}\triangleq\{1,...,K\}$ denote the set of user indices. At any time, each VR user is watching a rectangular part of the 360 VR video, referred to as FoV, the center of which is referred to as the viewing direction. A user may be interested in one FoV at sometime, and can freely switch to another FoV after a while.
Viewing direction (or FoV) prediction has been widely studied.
In this paper, we focus on transmission design for given prediction results. The proposed framework does not rely on any particular prediction method.

\subsection{Multi-Quality Tiled 360 VR Video}
We consider tiling to enable flexible transmission of necessary tiles so as to improve transmission efficiency of the 360 VR video.
Specifically, the 360 VR video with a frame rate of $R$ (in frames per second) is divided into $M\times N$ rectangular segments of the same size, referred to as tiles, where $M$ and $N$ represent the numbers of segments in each column and row, respectively.
Define $\mathcal{M}\triangleq\{1,...,M\}$ and $\mathcal{N}\triangleq\{1,...,N\}$. The $(m,n)$-th tile refers to the tile in the $m$-th row and the $n$-th column, for all $m\in\mathcal{M}$ and $n\in\mathcal{N}$.
Considering user heterogeneity (e.g., in cellular usage costs, display resolutions of devices, channel conditions, etc.), we pre-encode each tile into $L$ representations corresponding to $L$ quality levels using HEVC or H.264, as in Dynamic Adaptive Streaming over HTTP (DASH). Let $\mathcal{L}\triangleq\{1,...,L\}$ denote the set of quality levels. For all $l\in\mathcal{L}$, the $l$-th representation of each tile corresponds to the $l$-th lowest quality. For ease of exposition, assume that tiles with the same quality level have the same encoding rate.
The encoding rate of the $l$-th representation of a tile is denoted by $D_{l}$ (in bits/s). Note that $D_{1}<D_{2}<...<D_{L}$.
We study the system for the duration of the playback time of multiple groups of pictures (GOPs),\footnote{The duration of the playback time of one GOP is usually 0.5-1 second.} over which the FoV of each user does not change.
Let $r_{k}\in\mathcal{L}$ denote the quality requirement of user $k\in \mathcal{K}$ which is fixed within the considered duration.

To deal with possible prediction errors, for each user, the set of tiles that cover the predicted FoV plus a safe margin, denoted by $\mathcal{G}_{k}$, are delivered, as in~\cite{zhou2018clustile,DBLP:journals/corr/CheungLMT17,8428401,8478317,Xie:2017:IQV:3123266.3123291,He2018Joint,nguyen2019optimal,DBLP:journals/corr/abs-1901-02203}. The size of the safe margin can be chosen according to the prediction accuracy, so that for all $k\in \mathcal{K}$, $\mathcal{G}_{k}$ covers all tiles in the actual FoV of user $k$ with a very high probability that meets the QoE requirement.
Let $\mathcal{G}\triangleq\bigcup_{k\in\mathcal{K}}\mathcal{G}_{k}$ denote the set of indices of the tiles that need to be transmitted considering all $K$ users.
For all $\mathcal{S}\subseteq\mathcal{K}, \mathcal{S}\neq\varnothing$, let
\begin{align}
\mathcal{P}_{\mathcal{S}}\triangleq\left(\bigcap_{k\in\mathcal{S}}\mathcal{G}_{k}\right)\bigcap\left(\mathcal{G}-\bigcup_{k\in\mathcal{K}\setminus\mathcal{S}}\mathcal{G}_{k}\right)\nonumber
\end{align}
denote the set of indices of the tiles that need to be transmitted to all users in $\mathcal{S}$ and are not needed by any user in $\mathcal{K}\setminus\mathcal{S}$.\footnote{Note that for all $\mathcal{S}\subseteq\mathcal{K}, \mathcal{S}\neq\varnothing$ with $\mathcal{P}_{\mathcal{S}}\neq\varnothing$, the quality levels of the tiles in $\mathcal{P}_{\mathcal{S}}$ needed by the users in $\mathcal{S}$ may be different.}
Define
\begin{align}
\mathcal{P}\triangleq\left\{\mathcal{P}_{\mathcal{S}}\,|\,\mathcal{P}_{\mathcal{S}}\neq\varnothing, \mathcal{S}\subseteq\mathcal{K},\ \mathcal{S}\neq\varnothing\right\},\nonumber\\
\mathcal{I}\triangleq\left\{\mathcal{S}\,|\,\mathcal{P}_{\mathcal{S}}\neq\varnothing, \mathcal{S}\subseteq\mathcal{K}, \mathcal{S}\neq\varnothing\right\}.\nonumber
\end{align}
Then $\mathcal{P}$ forms a partition of $\mathcal{G}$ and $\mathcal{I}$ specifies the user sets corresponding to the partition.
In contrast with~\cite{Ahmadi:2017:AMS:3126686.3126743}, for all $\mathcal{S}\in\mathcal{I}$, we jointly consider the tiles in $\mathcal{P}_{\mathcal{S}}$ instead of treating them separately to significantly reduce computational complexity for optimal wireless streaming. In addition, for ease of implementation, we assume that the quality levels of all tiles in $\mathcal{P}_{\mathcal{S}}$ transmitted to or played by user $k$ are the same, for all $\mathcal{S}\in\mathcal{I},k\in\mathcal{S}$.

\begin{Exam}[Illustration of $\mathcal{P}$ and $\mathcal{I}$]\label{Exam1}
As illustrated in Fig. \ref{system0}, we consider $K=4$, $M=4$, $N=8$, $L=3$, $\mathcal{G}_{1}=\{(1,1),(2,1),(1,2),(2,2),(1,3),(2,3)\}$, $\mathcal{G}_{2}=\{(1,3),(2,3),(1,4),(2,4),(1,5),(2,5)\}$, $\mathcal{G}_{3}=\{(2,4),$
$(3,4),(2,5),(3,5),(2,6),(3,6)\},$
$\mathcal{G}_{4}=\{(3,5),(4,5),(3,6),$
$(4,6),(3,7),(4,7)\}$. Then, we have $\mathcal{P}_{\{1\}}=\{(1,1),(2,1),$
$(1,2),(2,2)\}$, $\mathcal{P}_{\{2\}}=\{(1,4),(1,5)\}$, $\mathcal{P}_{\{3\}}=\{(3,4),(2,6)\}$, $\mathcal{P}_{\{4\}}=\{(4,5),(4,6),(3,7),(4,7)\}$,
$\mathcal{P}_{\{1,2\}}=\{(1,3),(2,3)\}$, $\mathcal{P}_{\{2,3\}}=\{(2,4),(2,5)\}$, $\mathcal{P}_{\{3,4\}}=\{(3,5),(3,6)\}$,
$\mathcal{P}=\{\mathcal{P}_{\{1\}},\mathcal{P}_{\{2\}},\mathcal{P}_{\{3\}},\mathcal{P}_{\{4\}},\mathcal{P}_{\{1,2\}},\mathcal{P}_{\{2,3\}},\mathcal{P}_{\{3,4\}}\}$ and $\mathcal{I}=\{\{1\},$
$\{2\},\{3\},\{4\},\{1,2\},\{2,3\},\{3,4\}\}$.
\end{Exam}

\subsection{Transmission and Playback}
Let $x_{\mathcal{S},k}$ denote the playback quality level selection variable with respect to the tiles in $\mathcal{P}_{\mathcal{S}}$ to be played by user $k$, where
\begin{align}
&x_{\mathcal{S},k}\in \mathcal{L},\quad \mathcal{S}\in\mathcal{I},\ k\in\mathcal{S}.\label{0x0}
\end{align}
Denote $\mathbf{x}\triangleq(x_{\mathcal{S},k})_{\mathcal{S}\in\mathcal{I},k\in\mathcal{S}}$.
When neighboring tiles have very different qualities, there are obvious seams, which severely affect the user perceived quality~\cite{Yu:2015:CAR:2814347.2814348}.
To guarantee QoE, we consider two requirements for quality variation in an FoV, i.e., the absolute smoothness requirement and the relative smoothness requirement.

\begin{itemize}
  \item Under the absolute smoothness requirement, all tiles in a user's FoV must be played at the same quality level.
Considering the quality requirement of user $k$, we have:
\begin{align}
&x_{\mathcal{S},k}=r_{k},\quad \mathcal{S}\in\mathcal{I},\ k\in\mathcal{S}.\label{0xr2}
\end{align}
  \item Under the relative smoothness requirement, tiles in a user's FoV can be played at quality levels within a certain range, as a user usually cannot tell slight quality variation for tiles in his FoV~\cite{wang2014mixing}.
Considering the quality requirement of user $k$, we have:
\begin{align}
&r_{k}\leq x_{\mathcal{S},k}\leq r_{k}+\Delta,\quad \mathcal{S}\in\mathcal{I},\ k\in\mathcal{S},\label{0xr3}
\end{align}
where $\Delta$ represents the tolerance for quality variation in an FoV.
\end{itemize}

Comparing \eqref{0xr2} and \eqref{0xr3}, it is clear that the constraints in \eqref{0xr3} are less restrictive than the constraints in \eqref{0xr2}.
That is, the relative smoothness requirement is a relaxed version of the absolute smoothness requirement.

Let $y_{\mathcal{S},k,l}$ denote the transmission quality level selection variable with respect to quality level $l$ and the tiles in $\mathcal{P}_{\mathcal{S}}$ to be transmitted to user $k$, where
\begin{align}
&y_{\mathcal{S},k,l}\in\{0,1\},\quad \mathcal{S}\in\mathcal{I},\ k\in\mathcal{S},\ l\in\mathcal{L},\label{x0}\\
&\sum\nolimits_{l\in\mathcal{L}}y_{\mathcal{S},k,l}=1,\quad \mathcal{S}\in\mathcal{I},\ k\in\mathcal{S}.\label{x1}
\end{align}
Here, $y_{\mathcal{S},k,l}=1$ indicates that the quality level of all tiles in $\mathcal{P}_{\mathcal{S}}$ to be transmitted to user $k$ is $l$,
and $y_{\mathcal{S},k,l}=0$ otherwise.
Note that \eqref{x1} ensures that the server transmits only one representation of each tile in $\mathcal{P}_{\mathcal{S}}$ to user $k\in\mathcal{S}$.
Denote $\mathbf{y}\triangleq(y_{\mathcal{S},k,l})_{\mathcal{S}\in\mathcal{I},k\in\mathcal{S},l\in\mathcal{L}}$.
The quality level of all tiles in $\mathcal{P}_{\mathcal{S}}$ transmitted to user $k$ is $\sum_{l\in\mathcal{L}}ly_{\mathcal{S},k,l}$.
Transcoding refers to decoding a video source (that has already been encoded) into an intermediate uncompressed format and then re-encoding it into the target format. It can be used for bit rate, frame rate and resolution reduction, and hence can be used to perform video quality reduction.
We consider two video playback modes, i.e., the direct-playback mode (without user transcoding) and the transcode-playback mode (with user transcoding).\footnote{Many chips for mobile devices, such as Qualcomm's chips for mobile phones, Ambarella's chips for camcorders and Intel's chips for portable computer or tablet computer, can perform transcoding.}

\begin{itemize}
  \item In the direct-playback mode, the server has to transmit the $x_{\mathcal{S},k}$-th representations of the tiles in $\mathcal{P}_{\mathcal{S}}$ to user $k$, i.e.,
\begin{align}
\sum\nolimits_{l\in\mathcal{L}}ly_{\mathcal{S},k,l}= x_{\mathcal{S},k},\quad \mathcal{S}\in\mathcal{I},\ k\in \mathcal{S}.\label{0xiyk}
\end{align}
  \item In the transcode-playback mode, each user can convert a representation of a tile at a certain quality level to a representation at a lower quality level.
Hence, it only requires the server to transmit the tiles in $\mathcal{P}_{\mathcal{S}}$ at a quality level no smaller than $x_{\mathcal{S},k}$ to user $k$, i.e.,
\begin{align}
\sum\nolimits_{l\in\mathcal{L}}ly_{\mathcal{S},k,l}\geq x_{\mathcal{S},k},\quad \mathcal{S}\in\mathcal{I},\ k\in \mathcal{S}.\label{xiyk}
\end{align}
Note that user transcoding involves computation and consumes energy.
For ease of exposition, we assume that for each tile, reducing the quality level of a video frame by one has computation load $c$ (in CPU cycles).\footnote{This can be generalized without affecting the optimization framework.}
Let $f_{k}$ (in CPU cycles/s) represent the computing capability of user $k$.
Considering heterogeneous hardware conditions at different users, we allow $f_{k},k\in \mathcal{K}$ to be different.
Thus, the transcoding power at user $k$ for reducing the quality level of a tile by one is $P_{k}\triangleq\kappa_{k}cR(f_{k})^{2}$ (in Watt), where $\kappa_{k}$ is the energy coefficient depending on the chip architecture at user $k$~\cite{tran2018joint}.
Then, the total transcoding power at all users is
\begin{align}
\sum_{\mathcal{S}\in\mathcal{I}}\sum_{k\in\mathcal{S}}|\mathcal{P}_{\mathcal{S}}|P_{k}\left(\sum_{l\in\mathcal{L}}ly_{\mathcal{S},k,l}-x_{\mathcal{S},k}\right).\nonumber
\end{align}
\end{itemize}

Comparing \eqref{0xiyk} and \eqref{xiyk}, it is obvious that the constraints in \eqref{xiyk} are less restrictive than the constraints in \eqref{0xiyk}. In other words, the transcode-playback mode is more flexible than the direct-playback mode.

\begin{figure*}[t]
\normalsize{
\centering
\subfigure[Case without user transcoding and with the absolute smoothness requirement. $x_{\{1,2\},2}=1$, $x_{\{2,3\},2}=1$, $y_{\{1,2\},2,1}=y_{\{2,3\},2,1}=1$, $y_{\{1,2\},2,2}=y_{\{1,2\},2,3}=y_{\{2,3\},2,2}=y_{\{2,3\},2,3}=0$.]{
\begin{minipage}[b]{8.2cm}
\centering
\includegraphics[width=8.2cm]{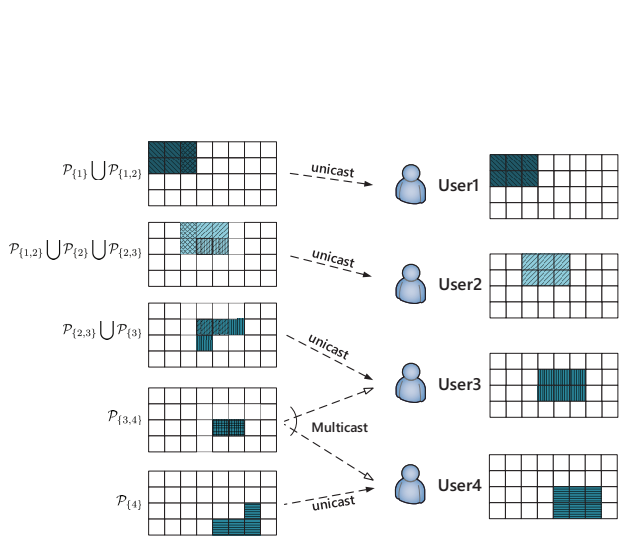}
\end{minipage}
}
\subfigure[Case without user transcoding and with the relative smoothness requirement. $x_{\{1,2\},2}=1$, $x_{\{2,3\},2}=2$, $y_{\{1,2\},2,1}=y_{\{2,3\},2,2}=1$, $y_{\{1,2\},2,2}=y_{\{1,2\},2,3}=y_{\{2,3\},2,1}=y_{\{2,3\},2,3}=0$.]{ 
\begin{minipage}[b]{8.2cm}
\centering
\includegraphics[width=8.2cm]{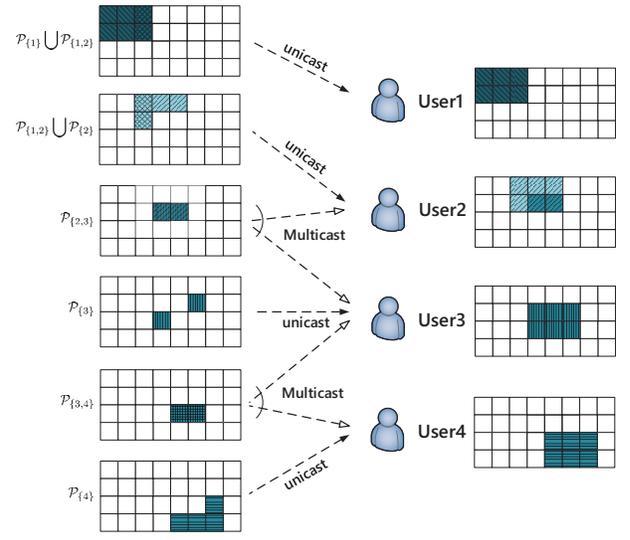}
\end{minipage}
}
\subfigure[Case with user transcoding and with the absolute smoothness requirement. $x_{\{1,2\},2}=1$, $x_{\{2,3\},2}=1$, $y_{\{1,2\},2,3}=y_{\{2,3\},2,1}=1$, $y_{\{1,2\},2,1}=y_{\{1,2\},2,2}=y_{\{2,3\},2,2}=y_{\{2,3\},2,3}=0$. ]{ 
\begin{minipage}[b]{8.2cm}
\centering
\includegraphics[width=8.2cm]{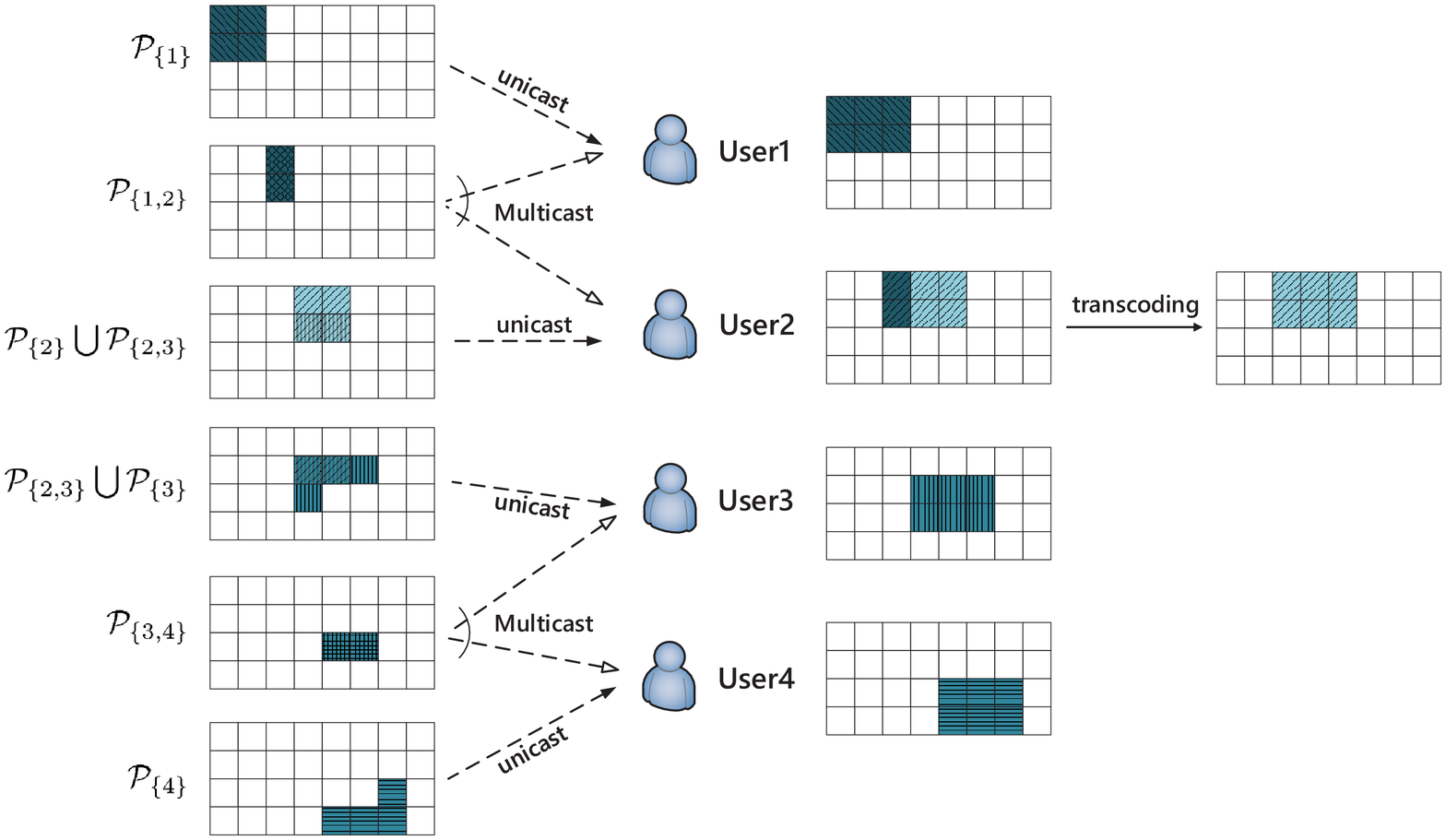}
\end{minipage}
}
\subfigure[Case with user transcoding and with the relative smoothness requirement. $x_{\{1,2\},2}=2$, $x_{\{2,3\},2}=2$, $y_{\{1,2\},2,3}=y_{\{2,3\},2,2}=1$, $y_{\{1,2\},2,1}=y_{\{1,2\},2,2}=y_{\{2,3\},2,1}=y_{\{2,3\},2,3}=0$. ]{ 
\begin{minipage}[b]{8.2cm}
\centering
\includegraphics[width=8.2cm]{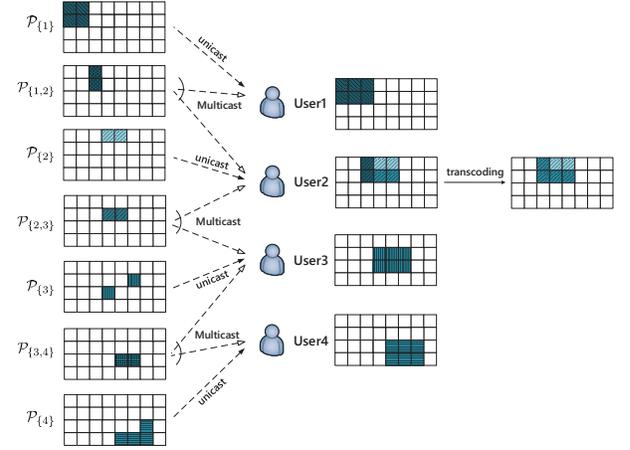}
\end{minipage}
}
\caption{Illustration of three types of multicast opportunities. The setup is the same as that in Fig. \ref{system0} and $\Delta=1$. In Fig.~\ref{system1}~(a),~(b),~(c) and~(d), $x_{\{1\},1}=3$, $x_{\{2\},2}=1$, $x_{\{3\},3}=2$, $x_{\{4\},4}=2$, $x_{\{1,2\},1}=3$, $x_{\{2,3\},3}=2$, $x_{\{3,4\},3}=2$, $x_{\{3,4\},4}=2$, $y_{\{1\},1,3}=y_{\{2\},2,1}=y_{\{3\},3,2}=y_{\{4\},4,2}=y_{\{1,2\},1,3}=y_{\{2,3\},3,2}=y_{\{3,4\},3,2}=y_{\{3,4\},4,2}=1$. $y_{\mathcal{S},k,l}=0$ for all other $\mathcal{S}\in\mathcal{I},k\in\mathcal{S},l\in\mathcal{L}$.}
\label{system1}
}
\end{figure*}

In summary, considering two smoothness requirements and two video playback modes, we have four cases, namely, the case without user transcoding and with the absolute smoothness requirement, the case without user transcoding and with the relative smoothness requirement, the case with user transcoding and with the absolute smoothness requirement, and the case with user transcoding and with the relative smoothness requirement, which possess different multicast opportunities.
In the following, we introduce three types of multicast opportunities in wireless streaming of the multi-quality tiled 360 VR video, as illustrated in Fig.~\ref{system1}. Consider any $\mathcal{S}\in\mathcal{I}$.

\begin{itemize}
  \item Natural multicast opportunities: If there exists $\mathcal{\underline{S}}\subseteq\mathcal{S}$ with $|\mathcal{\underline{S}}|\geq2$ such that $\mathcal{L}^{(\text{w/o,a})}_{\mathcal{\underline{S}}}\triangleq\bigcap_{k\in\mathcal{\underline{S}}}\{r_{k}\}\neq\varnothing$ (i.e., $r_{k}, k\in\mathcal{\underline{S}}$ are the same), then the server can multicast the $l_{\mathcal{\underline{S}}}$-th representations of the tiles in $\mathcal{P}_{\mathcal{S}}$ to simultaneously serve the users in $\mathcal{\underline{S}}$, where $l_{\mathcal{\underline{S}}}\in\mathcal{L}^{(\text{w/o,a})}_{\mathcal{\underline{S}}}$ (i.e., $l_{\mathcal{\underline{S}}}=r_{k}, k\in\mathcal{\underline{S}}$). We refer to this type of multicast opportunities as natural multicast opportunities.
  \item Relative smoothness-enabled multicast opportunities: If there exists $\mathcal{\underline{S}}\subseteq\mathcal{S}$ with $|\mathcal{\underline{S}}|\geq2$ such that $\mathcal{L}^{(\text{w/o,a})}_{\mathcal{\underline{S}}}=\varnothing$ and $\mathcal{L}^{(\text{r})}_{\mathcal{\underline{S}}}\triangleq\bigcap_{k\in\mathcal{\underline{S}}}\{r_{k},r_{k}+1,...,r_{k}+\Delta\}\neq\varnothing$, then the server can multicast the $l_{\mathcal{\underline{S}}}$-th representations of the tiles in $\mathcal{P}_{\mathcal{S}}$ to simultaneously serve the users in $\mathcal{\underline{S}}$, where $l_{\mathcal{\underline{S}}}\in\mathcal{L}^{(\text{r})}_{\mathcal{\underline{S}}}$. As for all $k\in\mathcal{\underline{S}}$, $r_{k}\leq l_{\mathcal{\underline{S}}}\leq r_{k}+\Delta$ is satisfied, user $k$ can directly play the received $l_{\mathcal{\underline{S}}}$-th representations of the tiles in $\mathcal{P}_{\mathcal{S}}$ under the relative smoothness requirement. We refer to this type of multicast opportunities as relative smoothness-enabled multicast opportunities.
  \item Transcoding-enabled multicast opportunities: If there exists $\mathcal{\underline{S}}\subseteq\mathcal{S}$ with $|\mathcal{\underline{S}}|\geq2$ such that
      $\mathcal{L}^{(\text{w/o,a})}_{\mathcal{\underline{S}}}=\varnothing$ and $\mathcal{L}^{(\text{w})}_{\mathcal{\underline{S}}}\triangleq\bigcap_{k\in\mathcal{\underline{S}}}\{r_{k},r_{k}+1,...,L\}\neq\varnothing$, then the server can multicast the $l_{\mathcal{\underline{S}}}$-th representations of the tiles in $\mathcal{P}_{\mathcal{S}}$ to simultaneously serve the users in $\mathcal{\underline{S}}$, where $l_{\mathcal{\underline{S}}}\in\mathcal{L}^{(\text{w})}_{\mathcal{\underline{S}}}$.
      Under the absolute smoothness requirement,
      for all $k\in\mathcal{\underline{S}}$ with $r_{k}=l_{\mathcal{\underline{S}}}$, user $k$ directly plays the received $l_{\mathcal{\underline{S}}}$-th representations of the tiles in $\mathcal{P}_{\mathcal{S}}$;
      for all $k\in\mathcal{\underline{S}}$ with $r_{k}<l_{\mathcal{\underline{S}}}$, user $k$ converts the $l_{\mathcal{\underline{S}}}$-th representations  of the tiles in $\mathcal{P}_{\mathcal{S}}$ to the $r_{k}$-th representations, and then plays them.
      Under the relative smoothness requirement,
      for all $k\in\mathcal{\underline{S}}$ with $r_{k}\leq l_{\mathcal{\underline{S}}}\leq r_{k}+\Delta$, user $k$ directly plays the received $l_{\mathcal{\underline{S}}}$-th representations of the tiles in $\mathcal{P}_{\mathcal{S}}$;
      for all $k\in\mathcal{\underline{S}}$ with $l_{\mathcal{\underline{S}}}> r_{k}+\Delta$, user $k$ converts the $l_{\mathcal{\underline{S}}}$-th representations of the tiles in $\mathcal{P}_{\mathcal{S}}$ to the ($r_{k}+\Delta)$-th representations, and then plays them.
      We refer to this type of multicast opportunities as transcoding-enabled multicast opportunities.
\end{itemize}

Note that natural multicast opportunities may exist in each of the four cases;
relative smoothness-enabled multicast opportunities may exist in each case with the relative smoothness requirement;
transcoding-enabled multicast opportunities may exist in each case with user transcoding.

\begin{Exam}[Illustration of Multicast Opportunities]\label{Exam2}
As shown in Fig.~\ref{system1},
consider the same setup as in Example \ref{Exam1} and suppose $\Delta=1$.
\begin{itemize}
  \item In each of the four cases, the server can multicast the second representations of the tiles in $\mathcal{P}_{\{3,4\}}$ to user 3 and user 4, by exploiting natural multicast opportunities. Both user 3 and user 4 directly play the received second representations of the tiles (as $r_{3}=r_{4}=2$).
  \item In the two cases with the relative smoothness requirement (as shown in Fig.~\ref{system1} (b) and Fig.~\ref{system1} (d)), the server can multicast the second representations of the tiles in $\mathcal{P}_{\{2,3\}}$ to user 2 and user 3, by exploiting relative smoothness-enabled multicast opportunities. Both user 2 and user 3 directly play the received second representations of the tiles (as $r_{2}+\Delta=1+1=2$ and $r_{3}=2$).
  \item In the two cases with user transcoding (as shown in Fig.~\ref{system1} (c) and Fig.~\ref{system1} (d)), the server can multicast the third representations of the tiles in $\mathcal{P}_{\{1,2\}}$ to user 1 and user 2, by exploiting transcoding-enabled multicast opportunities. In the case with user transcoding and with the absolute smoothness requirement (as shown in Fig.~\ref{system1} (c)), user 1 directly plays the received third representations of the tiles (as $r_{1}=3$), and user 2 converts the third representations to the first representations and then plays them (as $r_{2}=1$). In the case with user transcoding and with the relative smoothness requirement (as shown in Fig.~\ref{system1} (d)), user 1 directly plays the received third representations of the tiles (as $r_{1}=3$), and user 2 converts the third representations to the second representations and then plays them (as $r_{2}+\Delta=1+1=2$).
\end{itemize}

\end{Exam}

\subsection{TDMA Systems}
We consider a discrete narrowband system with time frame of duration $T$ (in seconds)\footnote{Note that $T$ is about 0.005-0.05 second.} and bandwidth $B$ (in Hz). Consider the block fading channel model, i.e., assume the channel of each user does not change within each time frame. For an arbitrary time frame, let $H_{k}\in\mathcal{H}$ denote the random channel state of user $k$, representing the power of the channel between user $k$ and the server, where $\mathcal{H}$ denotes the finite\footnote{Note that we consider a finite channel state space for tractability of optimization. In addition, note that due to limited accuracy for channel estimation (and channel feedback), the operational channel state space in practical systems is finite.} channel state space.\footnote{The overall attenuation over one path is the product of the attenuation factors due to the antenna pattern of the transmitter and the receiver, the nature of the reflector, as well as the path loss from the transmitting antenna to the receive antenna. The overall attenuations and propagation delays over different paths can be different and the channel power is a function of the overall attenuations and propagation delays over multiple paths~\cite{tse2005fundamentals}. In cellular systems, a base station can directly obtain a channel state by performing channel estimation.} Let $\mathbf{H}\triangleq(H_{k})_{k\in\mathcal{K}}\in\mathcal{H}^{K}$ denote the random system channel state in an arbitrary time frame, where $\mathcal{H}^{K}$ represents the finite system channel state space.
We assume that the server is aware of the system channel state $\mathbf{H}$ at each time frame. Suppose that the random system channel states over time frames are i.i.d.
The probability of the random system channel state $\mathbf{H}$ at each time frame being $\mathbf{h}\triangleq(h_{k})_{k\in\mathcal{K}}\in\mathcal{H}^{K}$ is given by $q_{\mathbf{H}}(\mathbf{h})\triangleq\mathrm{Pr}[\mathbf{H}=\mathbf{h}]$.

We consider TDMA for transmitting the set of tiles $\mathcal{P}_{\mathcal{S}}, \mathcal{S}\in\mathcal{I}$.
To save communications resource by making use of multicast opportunities, the server transmits one representation of each tile in $\mathcal{P}_{\mathcal{S}},\mathcal{S}\in\mathcal{I}$ at most once.
The time allocated to transmit the $l$-th representations of the tiles in $\mathcal{P}_{\mathcal{S}}$ under $\mathbf{h}$, denoted by $t_{\mathbf{h},\mathcal{S},l}$, satisfies:
\begin{align}
t_{\mathbf{h},\mathcal{S},l}\geq0,\quad\mathbf{h}\in\mathcal{H}^{K},\ \mathcal{S}\in\mathcal{I},\ l\in\mathcal{L}. \label{t1}
\end{align}
In addition, we have the following total time allocation constraint under $\mathbf{h}$:
\begin{align}
\sum\nolimits_{\mathcal{S}\in\mathcal{I}}\sum\nolimits_{l\in\mathcal{L}}t_{\mathbf{h},\mathcal{S},l}\leq T,\quad \mathbf{h}\in\mathcal{H}^{K}.\label{t2}
\end{align}
The power allocated to transmit the $l$-th representations of the tiles in $\mathcal{P}_{\mathcal{S}}$ under $\mathbf{h}$,
denoted by $p_{\mathbf{h},\mathcal{S},l}$, satisfies:
\begin{align}
p_{\mathbf{h},\mathcal{S},l}\geq0,\quad \mathbf{h}\in\mathcal{H}^{K},\ \mathcal{S}\in\mathcal{I},\ l\in\mathcal{L}.\label{pi}
\end{align}
The transmission energy per time frame under $\mathbf{h}$ at the server is
$\sum_{\mathcal{S}\in\mathcal{I}}\sum_{l\in\mathcal{L}}t_{\mathbf{h},\mathcal{S},l}p_{\mathbf{h},\mathcal{S},l}$,
and the average transmission energy per time frame is
\begin{align}
\mathbb{E}\left[\sum_{\mathcal{S}\in\mathcal{I}}\sum_{l\in\mathcal{L}}t_{\mathbf{H},\mathcal{S},l}p_{\mathbf{H},\mathcal{S},l}\right],\nonumber
\end{align}
where the expectation is taken over $\mathbf{H}\in \mathcal{H}^{K}$.
The maximum transmission rate of the $l$-th representations of the tiles in $\mathcal{P}_{\mathcal{S}}$ to user $k$ under $\mathbf{h}$ is given by
$\frac{B t_{\mathbf{h},\mathcal{S},l}}{T}\log_{2}\left(1+\frac{p_{\mathbf{h},\mathcal{S},l}h_{k}}{n_{0}}\right)$ (in bits/s)~\cite{tse2005fundamentals,xu2019optimal},
where $n_{0}$ is the power of the complex additive white Gaussian channel noise at each receiver.
To reduce the chance of stall (i.e., the chance that a playback buffer is empty) during the video playback at each user, we have the following successful transmission constraints~\cite{xu2019optimal}:
\begin{align}
|\mathcal{P}_{\mathcal{S}}|D_{l}y_{\mathcal{S},k,l}\leq \frac{B}{T}\mathbb{E}\left[t_{\mathbf{H},\mathcal{S},l}\log_{2}\left(1+\frac{p_{\mathbf{H},\mathcal{S},l}H_{k}}{n_{0}}\right)\right],\nonumber\\ \mathcal{S}\in\mathcal{I},\ k\in\mathcal{S},\ l\in\mathcal{L}, \label{xtp1}
\end{align}
where $|\mathcal{P}_{\mathcal{S}}|$ denotes the number of tiles in $\mathcal{P}_{\mathcal{S}}$.

\section{Optimal Wireless Streaming without User Transcoding}\label{section_3}
In this section, we consider optimal wireless streaming without user transcoding, and
minimize the average transmission energy under the absolute and relative smoothness requirements, separately, by exploiting respective multicast opportunities.

\subsection{Case without User Transcoding and with Absolute Smoothness Requirement}   \label{section_b1}
In this part, we consider the case without user transcoding and with the absolute smoothness requirement. Note that in this case, $\mathbf{x}$ is given by \eqref{0xr2}.
By \eqref{0xr2}, \eqref{x0}, \eqref{x1} and \eqref{0xiyk}, we have:
\begin{align}
y_{\mathcal{S},k,l}=
\begin{cases}
1,& l= r_{k}\\
0,& l\neq r_{k}
\end{cases},\quad \mathcal{S}\in\mathcal{I},\ k\in\mathcal{S}.\label{0xr21}
\end{align}
Hence, the successful transmission constraints in \eqref{xtp1} become:
\begin{align}
|\mathcal{P}_{\mathcal{S}}|D_{r_{k}}\leq \frac{B}{T}\mathbb{E}\left[t_{\mathbf{H},\mathcal{S},r_{k}}\log_{2}\left(1+\frac{p_{\mathbf{H},\mathcal{S},r_{k}}H_{k}}{n_{0}}\right)\right],\nonumber\\ \mathcal{S}\in\mathcal{I},\ k\in\mathcal{S}. \label{bxtp1}
\end{align}
We would like to optimize the transmission time allocation $\mathbf{t}\triangleq(\mathbf{t_{h}})_{\mathbf{h}\in\mathcal{H}^{K}}$ and transmission power allocation $\mathbf{p}\triangleq(\mathbf{p_{h}})_{\mathbf{h}\in\mathcal{H}^{K}}$ to minimize the average transmission energy subject to the transmission time allocation constraints in \eqref{t1}, \eqref{t2}, transmission power constraints in \eqref{pi}, and successful transmission constraints in \eqref{bxtp1}.

\begin{problem}[Without User Transcoding and with Absolute Smoothness Requirement]\label{bP1}
\begin{align}
E^{(\text{w/o,a})\star}\triangleq\min_{\mathbf{t},\mathbf{p}} \quad&\mathbb{E}\left[\sum_{\mathcal{S}\in\mathcal{I}}\sum_{l\in\mathcal{L}}t_{\mathbf{H},\mathcal{S},l}p_{\mathbf{H},\mathcal{S},l}\right]\nonumber\\
\text{s.t.}\quad&
\eqref{t1}, \eqref{t2}, \eqref{pi}, \eqref{bxtp1}.\nonumber
\end{align}
Let $\left(\mathbf{t}^{(\text{w/o,a})\star},\mathbf{p}^{(\text{w/o,a})\star}\right)$ denote an optimal solution of Problem~\ref{bP1}.
\end{problem}

Problem~\ref{bP1} is non-convex. In the following, we develop an algorithm to obtain an optimal solution of Problem~\ref{bP1}.
First, by a change of variables, i.e., using $e_{\mathbf{h},\mathcal{S},l}\triangleq t_{\mathbf{h},\mathcal{S},l}p_{\mathbf{h},\mathcal{S},l}$ (representing the transmission energy per time frame for the $l$-th representations of the tiles in $\mathcal{P}_{\mathcal{S}}$ under $\mathbf{h}$) instead of $p_{\mathbf{h},\mathcal{S},l}$, for all $h\in\mathcal{H}^{K},\mathcal{S}\in\mathcal{I},l\in\mathcal{L}$,
we can equivalently convert Problem~\ref{bP1} to the following convex problem.

\begin{problem}[Convex Formulation of Problem~\ref{bP1}]\label{bP2}
\begin{align}
E^{(\text{w/o,a})\star}\triangleq\min_{\mathbf{t},\mathbf{e}} &\quad\mathbb{E}\left[\sum_{\mathcal{S}\in\mathcal{I}}\sum_{l\in\mathcal{L}}e_{\mathbf{H},\mathcal{S},l}\right]\nonumber\\
\text{s.t.}
&\quad\eqref{t1}, \eqref{t2},\nonumber\\
&\quad e_{\mathbf{h},\mathcal{S},l}\geq0,\quad\mathbf{h}\in\mathcal{H}^{K},\mathcal{S}\in\mathcal{I},l\in\mathcal{L},\label{ei1}
\end{align}
\begin{align}
\quad|\mathcal{P}_{\mathcal{S}}|D_{r_{k}}\leq \frac{B}{T}\mathbb{E}\left[t_{\mathbf{H},\mathcal{S},r_{k}}\log_{2}\left(1+\frac{e_{\mathbf{H},\mathcal{S},r_{k}}H_{k}}{t_{\mathbf{H},\mathcal{S},r_{k}}n_{0}}\right)\right],\nonumber\\ \quad\mathcal{S}\in\mathcal{I},\ k\in\mathcal{S}. \label{bxte1}
\end{align}
Let $\left(\mathbf{t}^{(\text{w/o,a})\star},\mathbf{e}^{(\text{w/o,a})\star}\right)$ denote an optimal solution of Problem~\ref{bP2}.
\end{problem}

Due to the equivalence between Problem~\ref{bP1} and Problem~\ref{bP2}, we have:
\begin{align}
p^{(\text{w/o,a})\star}_{\mathbf{h},\mathcal{S},l}=e^{(\text{w/o,a})\star}_{\mathbf{h},\mathcal{S},l}/t^{(\text{w/o,a})\star}_{\mathbf{h},\mathcal{S},l},\mathbf{h}\in\mathcal{H}^{K},\mathcal{S}\in\mathcal{I},l\in\mathcal{L}.\nonumber
\end{align}
As Problem~\ref{bP2} is convex, an optimal solution of it can be obtained using standard convex optimization techniques~\cite{boyd2004convex}.
When $\mathcal{H}$ and $K$ are large, the numbers of variables and constraints in Problem~\ref{bP2} are huge, leading to prohibitively high computational complexity.
Divide variables $(\mathbf t,\mathbf e)$ into blocks $(\mathbf t_\mathbf{h},\mathbf e_\mathbf{h})$, $\mathbf{h}\in\mathcal H^K$, one for each $\mathbf{h}\in\mathcal H^K$. Note that the objective and the constraints in \eqref{t1}, \eqref{t2}, \eqref{ei1} of Problem~\ref{bP2} are block separable, and the constraints in $\eqref{bxte1}$ are the only coupling constraints that involve variables
from different blocks. In addition, note that Problem~\ref{bP2} is convex and strictly feasible, implying that Slater’s condition holds, and hence the duality gap is zero. Therefore, we can adopt partial dual decomposition and perform parallel computation to accelerate the speed for solving Problem~\ref{bP2}~\cite{boyd2007notes}.
Specifically, by relaxing the coupling constraints in $\eqref{bxte1}$, we can obtain a decomposable partial dual problem of Problem~\ref{bP2}, which shares the same optimal value as Problem~\ref{bP2}.

\begin{problem}[Partial Dual Decomposition of Problem~\ref{bP2}]\label{P31}
\begin{align}
\max_{\boldsymbol{\lambda}}\quad &D(\boldsymbol{\lambda})\label{P310}\\
\text{s.t.} \quad &\lambda_{\mathcal{S},k}\geq0,\quad \mathcal{S}\in\mathcal{I},\ k\in\mathcal{S},\nonumber
\end{align}
where $D(\boldsymbol{\lambda})\triangleq\sum\nolimits_{\mathbf{h}\in\mathcal{H}^{K}}q_{\mathbf{H}}(\mathbf{h})D_{\mathbf{h}}(\boldsymbol{\lambda})$ and
\begin{align}
D_{\mathbf{h}}(\boldsymbol{\lambda})\triangleq\nonumber\\
\min_{\mathbf{t_{h},e_{h}}} &\quad
\sum_{\mathcal{S}\in\mathcal{I}}\sum_{l\in\mathcal{L}}e_{\mathbf{h},\mathcal{S},l}-\sum_{\mathcal{S}\in\mathcal{I}}\sum_{k\in\mathcal{S}}\lambda_{\mathcal{S},k}\nonumber\\
&\quad \times\left(\frac{B}{T}t_{\mathbf{h},\mathcal{S},r_{k}}\log_{2}\left(1+\frac{e_{\mathbf{h},\mathcal{S},r_{k}}h_{k}}{t_{\mathbf{h},\mathcal{S},r_{k}}n_{0}}\right)-|\mathcal{P}_{\mathcal{S}}|D_{r_{k}}\right)\label{P33}\\
\text{s.t.}
&\quad t_{\mathbf{h},\mathcal{S},l}\geq0,\quad \mathcal{S}\in\mathcal{I},\ l\in\mathcal{L},\nonumber\\
&\quad\sum\nolimits_{\mathcal{S}\in\mathcal{I}}t_{\mathbf{h},\mathcal{S},l}\leq T,\quad l\in\mathcal{L},\nonumber\\
&\quad e_{\mathbf{h},\mathcal{S},l}\geq0,\quad \mathcal{S}\in\mathcal{I},\ l\in\mathcal{L}.\nonumber
\end{align}
Here, $\boldsymbol{\lambda}\triangleq(\lambda_{\mathcal{S},k})_{\mathcal{S}\in\mathcal{I},k\in\mathcal{S}}$, $\mathbf{t_{h}}\triangleq (t_{\mathbf{h},\mathcal{S},l})_{\mathcal{S}\in \mathcal{I},l\in\mathcal{L}}$ and $\mathbf{e_{h}}\triangleq (e_{\mathbf{h},\mathcal{S},l})_{\mathcal{S}\in \mathcal{I},l\in\mathcal{L}}$.
Let $\boldsymbol{\lambda}^{(\text{w/o,a})\star}$ denote an optimal solution of the optimization in~\eqref{P310} and let $(\mathbf{t}_{\mathbf{h}}^{(\text{w/o,a})\star}(\boldsymbol{\lambda}),\mathbf{e}_{\mathbf{h}}^{(\text{w/o,a})\star}(\boldsymbol{\lambda}))$ denote an optimal solution of the optimization in~\eqref{P33}, where $\mathbf{t}_{\mathbf{h}}^{(\text{w/o,a})\star}(\boldsymbol{\lambda})\triangleq(t^{(\text{w/o,a})\star}_{\mathbf{h},\mathcal{S},l}(\boldsymbol{\lambda}))_{\mathcal{S}\in\mathcal{I},l\in\mathcal{L}}$ and $\mathbf{e}_{\mathbf{h}}^{(\text{w/o,a})\star}(\boldsymbol{\lambda})\triangleq(e^{(\text{w/o,a})\star}_{\mathbf{h},\mathcal{S},l}(\boldsymbol{\lambda}))_{\mathcal{S}\in\mathcal{I},l\in\mathcal{L}}$.
\end{problem}

\begin{theorem}[Relationship between Problem~\ref{P31} and Problem~\ref{bP2}]\label{lemma1}
 $\mathbf{t}^{(\text{w/o,a})\star}=\mathbf{t}^{(\text{w/o,a})\star}(\boldsymbol{\lambda}^{(\text{w/o,a})\star})$ and $\mathbf{e}^{(\text{w/o,a})\star}=\mathbf{e}^{(\text{w/o,a})\star}(\boldsymbol{\lambda}^{(\text{w/o,a})\star})$, where $\mathbf{t}^{(\text{w/o,a})\star}(\boldsymbol{\lambda})\triangleq(\mathbf{t}_{\mathbf{h}}^{(\text{w/o,a})\star}(\boldsymbol{\lambda}))_{\mathbf{h}\in\mathcal{H}^{K}}$ and $\mathbf{e}^{(\text{w/o,a})\star}(\boldsymbol{\lambda})\triangleq(\mathbf{e}_{\mathbf{h}}^{(\text{w/o,a})\star}(\boldsymbol{\lambda}))_{\mathbf{h}\in\mathcal{H}^{K}}$.
\begin{IEEEproof}
Please refer to Appendix A.
\end{IEEEproof}
\end{theorem}

By Theorem~\ref{lemma1}, we can obtain an optimal solution of Problem~\ref{bP2} by solving Problem~\ref{P31}.
As the optimizations
 in~\eqref{P33} for all $\mathbf{h}\in\mathcal{H}^{K}$ are convex and can be solved in parallel using standard convex optimization techniques, we can compute $(\mathbf{t}^{(\text{w/o,a})\star}(\boldsymbol{\lambda}),\mathbf{e}^{(\text{w/o,a})\star}(\boldsymbol{\lambda}))$ efficiently. In addition, the optimization in~\eqref{P310} is convex and can be solved using the subgradient method~\cite{bertsekas1999nonlinear}.
Denote $\boldsymbol{\lambda}(n)\triangleq(\lambda_{\mathcal{S},k}(n))_{\mathcal{S}\in\mathcal{I},k\in\mathcal{S}}$. The details are summarized in Algorithm~\ref{alg0}. It has been shown in~\cite{bertsekas1999nonlinear} that, for all initial points $\boldsymbol{\lambda}(0)\succeq0$, $\boldsymbol{\lambda}(n)\rightarrow\boldsymbol{\lambda}^{(\text{w/o,a})\star}$, $(\mathbf{t}_{\mathbf{h}}^{(\text{w/o,a})\star}(\boldsymbol{\lambda}(n)))_{\mathbf{h}\in\mathcal{H}^{K}}$$\rightarrow$$\mathbf{t}^{(\text{w/o,a})\star}$ and $(\mathbf{e}_{\mathbf{h}}^{(\text{w/o,a})\star}(\boldsymbol{\lambda}(n)))_{\mathbf{h}\in\mathcal{H}^{K}}$$\rightarrow$$\mathbf{e}^{(\text{w/o,a})\star}$, as $n\rightarrow\infty$.

\begin{algorithm}[t]
\caption{Algorithm for Obtaining an Optimal Solution of Problem~\ref{bP1}}
\begin{footnotesize}
\begin{algorithmic}[1]\label{alg0}
\STATE Set $n=0$, and choose any $\boldsymbol{\lambda}(0)\succeq0$.
\REPEAT
\STATE Set $n=n+1$.
\STATE For all $\mathbf{h}\in\mathcal{H}^{K}$, obtain $(\mathbf{t}_{\mathbf{h}}^{(\text{w/o,a})\star}(\boldsymbol{\lambda}(n)),\mathbf{e}_{\mathbf{h}}^{(\text{w/o,a})\star}(\boldsymbol{\lambda}(n)))$ by solving the optimization in~\eqref{P33} using standard convex optimization techniques.
\STATE For all $\mathcal{S}\in\mathcal{I}$ and $k\in\mathcal{S}$, update $\lambda_{\mathcal{S},k}(n)$ according to $\lambda_{\mathcal{S},k}(n+1)=\max\{\lambda_{\mathcal{S},k}(n)+\eta_{\mathcal{S},k}(n)s_{\mathcal{S},k}(\boldsymbol{\lambda}(n)),0\}$,
where
\begin{align}
s_{\mathcal{S},k}(\boldsymbol{\lambda}(n))\triangleq |\mathcal{P}_{\mathcal{S}}|D_{r_{k}}-\frac{B}{T}\sum\nolimits_{\mathbf{h}\in\mathcal{H}^{K}}q_{\mathbf{H}}(\mathbf{h})t_{\mathbf{h},\mathcal{S},r_{k}}^{(\text{w/o,a})\star}(\boldsymbol{\lambda}(n))\nonumber\\
\times\log_{2}\left(1+\frac{e_{\mathbf{h},\mathcal{S},r_{k}}^{(\text{w/o,a})\star}(\boldsymbol{\lambda}(n))h_{k}}{t_{\mathbf{h},\mathcal{S},r_{k}}^{(\text{w/o,a})\star}(\boldsymbol{\lambda}(n))n_{0}}\right),\nonumber
\end{align} and $\{\eta_{\mathcal{S},k}(n)\}$ is a step size sequence satisfying
$\eta_{\mathcal{S},k}(n)>0,\; \sum_{n=0}^{\infty}\eta_{\mathcal{S},k}(n)=\infty,\; \sum_{n=0}^{\infty}\eta_{\mathcal{S},k}^{2}(n)<\infty$.
\UNTIL{some convergence criterion is met.}
\end{algorithmic}
\end{footnotesize}
\end{algorithm}

\subsection{Case without User Transcoding and with Relative Smoothness Requirement}  \label{section_b2}
In this part, we consider the case without user transcoding and with the relative smoothness requirement.
By \eqref{0xiyk}, $\mathbf{x}$ can be determined by $\mathbf{y}$, and the constraints in \eqref{0xr3} become:
\begin{align}
r_{k}\leq\sum\nolimits_{l\in\mathcal{L}}ly_{\mathcal{S},k,l}\leq r_{k}+\Delta,\quad \mathcal{S}\in\mathcal{I},\ k\in \mathcal{S}.\label{xiyk1}
\end{align}
We would like to optimize the transmission quality level selection $\mathbf{y}$, transmission time allocation $\mathbf{t}$ and transmission power allocation $\mathbf{p}$ to minimize the average transmission energy
subject to the transmission quality level selection constraints in \eqref{x0}, \eqref{x1}, \eqref{xiyk1}, transmission time allocation constraints in \eqref{t1}, \eqref{t2}, transmission power constraints in \eqref{pi}, and
successful transmission constraints in \eqref{xtp1}.
\begin{problem}[Without User Transcoding and with Relative Smoothness Requirement]\label{sP1}
\begin{align}
{E}^{(\text{w/o,r})\star}\triangleq\min_{\mathbf{y},\mathbf{t},\mathbf{p}} \quad&\mathbb{E}\left[\sum_{\mathcal{S}\in\mathcal{I}}\sum_{l\in\mathcal{L}}t_{\mathbf{H},\mathcal{S},l}p_{\mathbf{H},\mathcal{S},l}\right]\nonumber\\
\text{s.t.}\quad&
\eqref{x0}, \eqref{x1}, \eqref{t1}, \eqref{t2}, \eqref{pi}, \eqref{xtp1}, \eqref{xiyk1}.\nonumber
\end{align}
Let $(\mathbf{{y}}^{(\text{w/o,r})\star},\mathbf{{t}}^{(\text{w/o,r})\star},\mathbf{{p}}^{(\text{w/o,r})\star})$ denote an optimal solution of Problem~\ref{sP1}.
\end{problem}

Problem~\ref{sP1} is a challenging mixed discrete-continuous optimization problem.
In the following, we develop an algorithm to obtain a suboptimal solution of Problem~\ref{sP1} using convex concave procedure.

First, we convert Problem~\ref{sP1} to a penalized DC programming. Similarly, by a change of variables, we use $\mathbf{e}$ instead of $\mathbf{p}$.
Besides, we equivalently convert the discrete constraints in \eqref{x0} to the following continuous constraints:
\begin{align}
&0\leq y_{\mathcal{S},k,l}\leq1, \quad \mathcal{S}\in\mathcal{I},\ k\in\mathcal{S},\ l\in\mathcal{L},\label{x2}\\
&y_{\mathcal{S},k,l}(1-y_{\mathcal{S},k,l})\leq 0, \quad \mathcal{S}\in\mathcal{I},\ k\in\mathcal{S},\ l\in\mathcal{L}.\label{x3}
\end{align}
By disregarding the constraints in \eqref{x3} and adding to the objective function a penalty for violating them, we can convert Problem~\ref{sP1} to the following problem.
\begin{problem}[Penalized DC Programming of Problem~\ref{sP1}]\label{sP2}
\begin{align}
\min_{\mathbf{y},\mathbf{t},\mathbf{e}} \quad
&\mathbb{E}\left[\sum_{\mathcal{S}\in\mathcal{I}}\sum_{l\in\mathcal{L}}e_{\mathbf{H},\mathcal{S},l}\right]+\rho P(\mathbf{y})\nonumber\\
\text{s.t.} \quad
&\eqref{x1}, \eqref{t1}, \eqref{t2},\eqref{ei1}, \eqref{xiyk1}, \eqref{x2},\nonumber\\
&|\mathcal{P}_{\mathcal{S}}|D_{l}y_{\mathcal{S},k,l}\leq \frac{B}{T}\mathbb{E}\left[t_{\mathbf{H},\mathcal{S},l}\log_{2}\left(1+\frac{e_{\mathbf{H},\mathcal{S},l}H_{k}}{t_{\mathbf{H},\mathcal{S},l}n_{0}}\right)\right],\nonumber\\ &\qquad\qquad\qquad\qquad\qquad\mathcal{S}\in\mathcal{I},\ k\in\mathcal{S},\ l\in\mathcal{L},\label{xte1}
\end{align}
where the penalty parameter $\rho>0$ and the penalty function $P(\mathbf{y})$ is given by
\begin{align}
P(\mathbf{y})\triangleq\sum\limits_{\mathcal{S}\in\mathcal{I}}\sum\limits_{k\in\mathcal{S}}\sum\limits_{l\in\mathcal{L}}y_{\mathcal{S},k,l}(1-y_{\mathcal{S},k,l}).\label{penaltyf}
\end{align}
\end{problem}

Note that the objective function of Problem~\ref{sP2} can be viewed as a difference of two convex functions and the feasible set of Problem~\ref{sP2} is convex. Thus, Problem~\ref{sP2} can be viewed as a penalized DC programming of Problem~\ref{sP1}.
An optimal solution of Problem~\ref{sP2} with zero penalty is also optimal for Problem~\ref{sP1}.

Next, we solve Problem~\ref{sP2} instead of Problem~\ref{sP1} by using convex concave procedure \cite{Lipp2016}. The main idea is to iteratively solve a sequence of convex approximations of Problem~\ref{sP2}, each of which is obtained by linearizing the penalty function $P(\mathbf{y})$ in \eqref{penaltyf}. Specifically, the convex approximation of Problem~\ref{sP2} at the $j$-th iteration is given below.

\begin{problem}[Convex Approximation of Problem~\ref{sP2} at $j$-th Iteration]\label{sP3}
\begin{align}
(\mathbf{{y}}^{(j)},\mathbf{{t}}^{(j)},\mathbf{{e}}^{(j)})\triangleq\nonumber\\
\mathop{\arg\min}_{\mathbf{y},\mathbf{t},\mathbf{e}} \ &\mathbb{E}\left[\sum_{\mathcal{S}\in\mathcal{I}}\sum_{l\in\mathcal{L}}e_{\mathbf{H},\mathcal{S},l}\right]+\rho \widehat{P}(\mathbf{y},\mathbf{{y}}^{(j-1)})\nonumber\\
\text{s.t.} \quad&\eqref{x1}, \eqref{t1}, \eqref{t2}, \eqref{ei1}, \eqref{xiyk1}, \eqref{x2}, \eqref{xte1},\nonumber
\end{align}
where
\begin{align}
&\hat{P}(\mathbf{y},\mathbf{{y}}^{(j-1)})\triangleq P(\mathbf{{y}}^{(j-1)})+\nabla P(\mathbf{{y}}^{(j-1)})^T (\mathbf{y}-\mathbf{{y}}^{(j-1)})\nonumber\\
&=\sum_{\mathcal{S}\in\mathcal{I}}\sum_{k\in\mathcal{S}}\sum_{l\in\mathcal{L}} \big(\big(1-2{y}_{\mathcal{S},k,l}^{(j-1)}\big)y_{\mathcal{S},k,l}+\big({y}_{\mathcal{S},k,l}^{(j-1)}\big)^2\big),\nonumber
\end{align}
and $\mathbf{{y}}^{(j-1)}\triangleq \left({y}_{\mathcal{S},k,l}^{(j-1)}\right)_{\mathcal{S}\in\mathcal{I},k\in\mathcal{S},l\in\mathcal{L}}$ denotes an optimal solution of Problem~\ref{sP3} at the $(j-1)$-th iteration.
\end{problem}

Problem~\ref{sP3} is a convex optimization problem and can be solved using standard convex optimization techniques or
partial dual decomposition and parallel computation as in Section~\ref{section_b1}.
It is known that the sequence $\{(\mathbf{{y}}^{(j)},\mathbf{{t}}^{(j)},\mathbf{{e}}^{(j)})\}$ generated by convex concave procedure is convergent, and the
limit point of $\{(\mathbf{{y}}^{(j)},\mathbf{{t}}^{(j)},\mathbf{{e}}^{(j)})\}$ is a stationary point of Problem~\ref{sP2}. We can run convex concave procedure multiple times, each with a random initial feasible point of Problem~\ref{sP2},
and select the stationary point with the minimum energy among those with zero penalty, denoted by $(\mathbf{{y}}^{(\text{w/o,r})\dag},\mathbf{{t}}^{(\text{w/o,r})\dag},\mathbf{{e}}^{(\text{w/o,r})\dag})$.
Due to the equivalence between Problem~\ref{sP1} and Problem~\ref{sP2}, for sufficiently large $\rho$, $\left(\mathbf{{y}}^{(\text{w/o,r})\dag},\mathbf{{t}}^{(\text{w/o,r})\dag},\mathbf{{p}}^{(\text{w/o,r})\dag}\right)$ can be treated as a suboptimal solution of Problem~\ref{sP1}, where
$\mathbf{{p}}^{(\text{w/o,r})\dag}\triangleq({p}^{(\text{w/o,r})\dag}_{\mathbf{h},\mathcal{S},l})_{\mathbf{h}\in\mathcal{H}^{K},\mathcal{S}\in\mathcal{I},l\in\mathcal{L}}$ with ${p}^{(\text{w/o,r})\dag}_{\mathbf{h},\mathcal{S},l}={e}^{(\text{w/o,r})\dag}_{\mathbf{h},\mathcal{S},l}/{t}^{(\text{w/o,r})\dag}_{\mathbf{h},\mathcal{S},l}$.
The details for obtaining a suboptimal solution of Problem~\ref{sP1} using convex concave procedure are summarized in Algorithm~\ref{alg2}.

\begin{algorithm}[t]
\caption{Algorithm for Obtaining a Suboptimal Solution of Problem~\ref{sP1}}
\begin{footnotesize}
\textbf{Input:} $c\geq1$.
\begin{algorithmic}[1]\label{alg2}
\STATE Set $E=+\infty$.
\WHILE {$c>0$}
\STATE Find a random feasible point of Problem~\ref{sP1} as the initial point $(\mathbf{{y}}^{(0)},\mathbf{{t}}^{(0)},\mathbf{{e}}^{(0)})$, choose a sufficiently large $\rho$, and set $j=0$.
\REPEAT
\STATE Set $j=j+1$.
\STATE Obtain $(\mathbf{{y}}^{(j)},\mathbf{{t}}^{(j)},\mathbf{{e}}^{(j)})$ by solving Problem~\ref{sP3} using standard convex optimization techniques or partial dual decomposition and parallel computation.
\UNTIL{some convergence criterion is met.}
\IF{${P}(\mathbf{{y}}^{(j)})=0$}
    \STATE Set $c=c-1$.
    \IF{$\;\mathbb{E}[\sum_{\mathcal{S}\in\mathcal{I}}\sum_{l\in\mathcal{L}}{e}^{(j)}_{\mathbf{H},\mathcal{S},l}]<E$}
    \STATE Set $E=\mathbb{E}[\sum_{\mathcal{S}\in\mathcal{I}}\sum_{l\in\mathcal{L}}{e}^{(j)}_{\mathbf{H},\mathcal{S},l}]$,
    $\mathbf{{y}}^{(\text{w/o,r})\dag}$$=$$\mathbf{{y}}^{(j)}$, $\mathbf{{t}}^{(\text{w/o,r})\dag}$$=$$\mathbf{{t}}^{(j)}$ and $\mathbf{{p}}^{(\text{w/o,r})\dag}$$=$$({p}^{(\text{w/o,r})\dag}_{\mathbf{h},\mathcal{S},l})_{\mathbf{h}\in\mathcal{H}^{K},\mathcal{S}\in\mathcal{I},l\in\mathcal{L}}$ with ${p}^{(\text{w/o,r})\dag}_{\mathbf{h},\mathcal{S},l}={e}^{(j)}_{\mathbf{h},\mathcal{S},l}/{t}^{(j)}_{\mathbf{h},\mathcal{S},l}$.
    \ENDIF
\ENDIF
\ENDWHILE
\end{algorithmic}
\end{footnotesize}
\end{algorithm}

\section{Optimal Wireless Streaming With User Transcoding}\label{section_4}
In this section, we consider optimal wireless streaming with user transcoding, and minimize the weighted sum of the average transmission energy and the transcoding energy per time frame, i.e.,
\begin{align}
&\mathbb{E}\left[\sum_{\mathcal{S}\in\mathcal{I}}\sum_{l\in\mathcal{L}}t_{\mathbf{H},\mathcal{S},l}p_{\mathbf{H},\mathcal{S},l}\right]\nonumber\\
&+\beta\sum_{\mathcal{S}\in\mathcal{I}}\sum_{k\in\mathcal{S}}|\mathcal{P}_{\mathcal{S}}|P_{k}T\left(\sum_{l\in\mathcal{L}}ly_{\mathcal{S},k,l}-x_{\mathcal{S},k}\right),\label{energysum}
\end{align}
under the absolute and relative smoothness requirements separately, by exploiting respective multicast opportunities.
Here, $\beta\geq1$ with $\beta>1$ meaning imposing a higher cost on the energy consumption for user devices due to their limited battery powers.

\subsection{Case with User Transcoding and with Absolute Smoothness Requirement}  \label{section_c1}
In this part, we consider the case with user transcoding and with the absolute smoothness requirement.
Note that in this case, $\mathbf{x}$ is given by \eqref{0xr2}.
By \eqref{0xr2}, the constraints in \eqref{xiyk} become:
\begin{align}
\sum\nolimits_{l\in\mathcal{L}}ly_{\mathcal{S},k,l}\geq r_{k},\quad \mathcal{S}\in\mathcal{I},\ k\in \mathcal{S},\label{xiyk01}
\end{align}
and the weighted sum average energy per time frame in \eqref{energysum} becomes:
\begin{align}
&\mathbb{E}\left[\sum_{\mathcal{S}\in\mathcal{I}}\sum_{l\in\mathcal{L}}t_{\mathbf{H},\mathcal{S},l}p_{\mathbf{H},\mathcal{S},l}\right]\nonumber\\
&+\beta\sum_{\mathcal{S}\in\mathcal{I}}\sum_{k\in\mathcal{S}}|\mathcal{P}_{\mathcal{S}}|P_{k}T\left(\sum_{l\in\mathcal{L}}ly_{\mathcal{S},k,l}-r_{k}\right).\nonumber
\end{align}
We would like to optimize the transmission quality level selection  $\mathbf{y}$, transmission time allocation $\mathbf{t}$ and transmission power allocation $\mathbf{p}$ to minimize the weighted sum average energy subject to the transmission quality level selection constraints in \eqref{x0}, \eqref{x1}, \eqref{xiyk01}, transmission time allocation constraints in \eqref{t1}, \eqref{t2}, transmission power constraints in \eqref{pi}, and
successful transmission constraints in \eqref{xtp1}.

\begin{problem}[With User Transcoding and with Absolute Smoothness Requirement]\label{P1}
\begin{align}
{E}^{(\text{w,a})\star}\triangleq\min_{\mathbf{y},\mathbf{t},\mathbf{p}} \quad&\mathbb{E}\left[\sum_{\mathcal{S}\in\mathcal{I}}\sum_{l\in\mathcal{L}}t_{\mathbf{H},\mathcal{S},l}p_{\mathbf{H},\mathcal{S},l}\right]\nonumber\\
&+\beta\sum_{\mathcal{S}\in\mathcal{I}}\sum_{k\in\mathcal{S}}|\mathcal{P}_{\mathcal{S}}|P_{k}T\left(\sum_{l\in\mathcal{L}}ly_{\mathcal{S},k,l}-r_{k}\right)\nonumber\\
\text{s.t.}\quad&
\eqref{x0}, \eqref{x1}, \eqref{t1}, \eqref{t2}, \eqref{pi}, \eqref{xtp1}, \eqref{xiyk01}.\nonumber
\end{align}
Let $(\mathbf{{y}}^{(\text{w,a})\star},\mathbf{{t}}^{(\text{w,a})\star},\mathbf{{p}}^{(\text{w,a})\star})$ denote an optimal solution of Problem~\ref{P1}.
\end{problem}

Similar to Problem~\ref{sP1}, Problem~\ref{P1} is a challenging mixed discrete-continuous optimization problem. We can obtain a suboptimal solution of it using an algorithm similar to Algorithm~\ref{alg2}.
Specifically, by a change of variables, we use $\mathbf{e}$ instead of $\mathbf{p}$. In addition, we equivalently convert the discrete constraints in \eqref{x0} to the continuous constraints in \eqref{x2} and \eqref{x3}. By disregarding the constraints in \eqref{x3} and adding the penalty function $P(\mathbf{y})$ in \eqref{penaltyf} to the objective function of Problem~\ref{P1}, we can convert Problem~\ref{P1} to
a penalized DC programming. Then, we can obtain a suboptimal solution of the penalized DC programming by using convex concave procedure.

\subsection{Case with User Transcoding and with Relative Smoothness Requirement}  \label{section_c2}
We would like to optimize the playback quality level selection $\mathbf{x}$, transmission quality level selection $\mathbf{y}$, transmission time allocation $\mathbf{t}$ and transmission power allocation $\mathbf{p}$ to minimize the weighted sum average energy
subject to the quality level selection constraints in \eqref{0x0}, \eqref{0xr3}, \eqref{x0}, \eqref{x1}, \eqref{xiyk}, transmission time allocation constraints in \eqref{t1}, \eqref{t2}, transmission power constraints in \eqref{pi}, and
successful transmission constraints in \eqref{xtp1}.

\begin{problem}[With User Transcoding and with Relative Smoothness Requirement]\label{0ssP1}
\begin{align}
{E}^{(\text{w,r})\star}\triangleq\min_{\mathbf{x},\mathbf{y},\mathbf{t},\mathbf{p}} &\quad\mathbb{E}\left[\sum_{\mathcal{S}\in\mathcal{I}}\sum_{l\in\mathcal{L}}t_{\mathbf{H},\mathcal{S},l}p_{\mathbf{H},\mathcal{S},l}\right]\nonumber\\
&+\beta\sum_{\mathcal{S}\in\mathcal{I}}\sum_{k\in\mathcal{S}}|\mathcal{P}_{\mathcal{S}}|P_{k}T\left(\sum_{l\in\mathcal{L}}ly_{\mathcal{S},k,l}-x_{\mathcal{S},k}\right)\nonumber\\
\text{s.t.}&\quad
\eqref{0x0}, \eqref{0xr3}, \eqref{x0}, \eqref{x1}, \eqref{xiyk}, \eqref{t1}, \eqref{t2}, \eqref{pi}, \eqref{xtp1}.\nonumber
\end{align}
Let $(\mathbf{{x}}^{(\text{w,r})\star},\mathbf{{y}}^{(\text{w,r})\star},\mathbf{{t}}^{(\text{w,r})\star},\mathbf{{p}}^{(\text{w,r})\star})$ denote an optimal solution of Problem~\ref{0ssP1}, where
$\mathbf{{x}}^{(\text{w,r})\star}\triangleq({x}^{(\text{w,r})\star}_{\mathcal{S},k})_{\mathcal{S}\in\mathcal{I},k\in\mathcal{S}}$, $\mathbf{{y}}^{(\text{w,r})\star}\triangleq({y}^{(\text{w,r})\star}_{\mathcal{S},k,l})_{\mathcal{S}\in\mathcal{I},k\in\mathcal{S},l\in\mathcal{L}}$, $\mathbf{{t}}^{(\text{w,r})\star}\triangleq({t}^{(\text{w,r})\star}_{\mathbf{h},\mathcal{S},l})_{\mathbf{h}\in\mathcal{H}^{K},\mathcal{S}\in\mathcal{I},l\in\mathcal{L}}$ and $\mathbf{{p}}^{(\text{w,r})\star}\triangleq({p}^{(\text{w,r})\star}_{\mathbf{h},\mathcal{S},l})_{\mathbf{h}\in\mathcal{H}^{K},\mathcal{S}\in\mathcal{I},l\in\mathcal{L}}$.
\end{problem}

Problem~\ref{0ssP1} is also a challenging mixed discrete-continuous optimization problem.
First, we analyze optimality properties of Problem~\ref{0ssP1} to reduce computational complexity for solving Problem~\ref{0ssP1}.
\begin{theorem}[Optimality Properties of Problem~\ref{0ssP1}]\label{lemma2}
The optimal solution of Problem~\ref{0ssP1} satisfies:
\begin{align}
{x}^{(\text{w,r})\star}_{\mathcal{S},k}=\min\left\{{r}^{(\text{w,r})}_{k}+\Delta, \sum_{l\in\mathcal{L}}l{y}^{(\text{w,r})\star}_{\mathcal{S},k,l}\right\},\quad\mathcal{S}\in\mathcal{I},k\in\mathcal{S}.\label{lem4}
\end{align}
\begin{IEEEproof}
Please refer to Appendix B.
\end{IEEEproof}
\end{theorem}

By Theorem~\ref{lemma2}, we can eliminate $\mathbf{x}$ in Problem~\ref{0ssP1} without loss of optimality. Thus, we can equivalently convert Problem~\ref{0ssP1} to the following problem.
\begin{problem}[Equivalent Problem of Problem~\ref{0ssP1}]\label{ssP1}
\begin{align}
{E}^{(\text{w,r})\star}\triangleq\min_{\mathbf{y},\mathbf{t},\mathbf{p}} \quad&\mathbb{E}\left[\sum_{\mathcal{S}\in\mathcal{I}}\sum_{l\in\mathcal{L}}t_{\mathbf{H},\mathcal{S},l}p_{\mathbf{H},\mathcal{S},l}\right]+\beta\sum_{\mathcal{S}\in\mathcal{I}}\sum_{k\in\mathcal{S}}|\mathcal{P}_{\mathcal{S}}|\nonumber\\
&\times P_{k}T\max\left\{\sum_{l\in\mathcal{L}}ly_{\mathcal{S},k,l}-(r_{k}+\Delta),0\right\}\nonumber\\
\text{s.t.} \quad
&\eqref{x0},\eqref{x1},\eqref{t1},\eqref{t2},\eqref{pi}, \eqref{xtp1},\eqref{xiyk01}.\nonumber
\end{align}
\end{problem}

As $\max\left\{\sum_{l\in\mathcal{L}}ly_{\mathcal{S},k,l}-(r_{k}+\Delta),0\right\}$ is convex with respect to $y_{\mathcal{S},k,l},l\in\mathcal{L}$, the structure of Problem~\ref{ssP1} is the same as that of Problem~\ref{P1}. Therefore, similarly, we
can convert Problem~\ref{ssP1} to a penalized DC programming and obtain a suboptimal solution using convex concave procedure.

\begin{figure}[t]
\normalsize{
\centering
\subfigure{
\begin{minipage}{6cm}
\centering
\includegraphics[width=6cm]{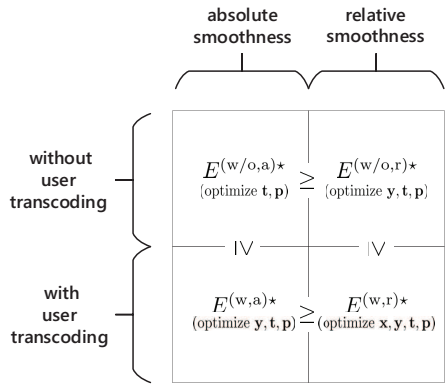}
\end{minipage}
}
\caption{Comparison of optimal designs in the four cases.}
\label{comp}
}
\end{figure}

\section{Discussion}
\subsection{Comparison of Optimal Designs in Four Cases}
In Section~\ref{section_3} and Section~\ref{section_4}, we have investigated the optimal wireless streaming in the four cases. As illustrated in Fig.~\ref{comp}, besides the optimal transmission time and power allocation considered in all four cases, the optimal playback quality level selection is optimized in the case with user transcoding and with the relative smoothness requirement, and the optimal transmission quality level selection is optimized in the three cases with user transcoding or with the relative smoothness requirement. As the optimization range increases, the optimal value reduces, as illustrated in Fig.~\ref{comp} and summarized in the following theorem.

\begin{theorem}[Comparison of Problem~\ref{bP1}, Problem~\ref{sP1}, Problem~\ref{P1} and Problem~\ref{0ssP1}]\label{compare0}
${E}^{(\text{w/o,r})\star} \leq E^{(\text{w/o,a})\star}$, ${E}^{(\text{w,r})\star}\leq{E}^{(\text{w,a})\star}$,
${E}^{(\text{w,a})\star}\leq E^{(\text{w/o,a})\star}$ and ${E}^{(\text{w,r})\star}\leq{E}^{(\text{w/o,r})\star}$,
where $E^{(\text{w/o,a})\star}$, ${E}^{(\text{w/o,r})\star}$, ${E}^{(\text{w,a})\star}$ and ${E}^{(\text{w,r})\star}$ are the optimal values of Problem~\ref{bP1}, Problem~\ref{sP1}, Problem~\ref{P1} and Problem~\ref{0ssP1}, respectively.
\begin{IEEEproof}
Please refer to Appendix C.
\end{IEEEproof}
\end{theorem}

\subsection{Extension}
In this paper, we adopt the first transmission scheme (which is illustrated in Section~\ref{section_1}) to demonstrate the key ideas. The proposed solution framework can be easily extended to the second transmission scheme or the combined transmission scheme (which are illustrated in Section~\ref{section_1}). In particular, additionally, the first representations of the tiles in $\mathcal{\bar{G}}_k\triangleq \mathcal{M}\times \mathcal{N}-\mathcal{G}_k $ are delivered to user $k$, for all $k\in\mathcal{K}$. Analogously, we can derive a partition of $\mathcal{\bar{G}}\triangleq\bigcup_{k\in\mathcal{K}}\mathcal{\bar{G}}_{k}$ and the corresponding user sets, and introduce the respective optimization variables. The corresponding playback quality level selection variables and transmission quality level selection variables can be determined as in the case without user transcoding and with the absolute smoothness requirement. The objective functions of Problem~\ref{bP1}, Problem~\ref{sP1}, Problem~\ref{P1} and Problem~\ref{0ssP1} in the four cases as well as the transmission time allocation constraints in \eqref{t1}, \eqref{t2}, the transmission power constraint in \eqref{pi} and the successful transmission constraints in \eqref{xtp1} can be modified to incorporate the additional transmission for the first representations of the sets of tiles in the partition of $\mathcal{\bar{G}}$. Therefore, for the second transmission scheme or the combined transmission scheme, we can formulate the optimization problems in the four cases which are similar to Problem~\ref{bP1}, Problem~\ref{sP1}, Problem~\ref{P1} and Problem~\ref{0ssP1}, and solve them using the proposed methods in Section~\ref{section_3} and Section~\ref{section_4}.

\section{Numerical Results}\label{section_5}
\begin{figure}[t]
\normalsize{
\centering
\subfigure{
\begin{minipage}{6cm}
\centering
\includegraphics[width=6cm]{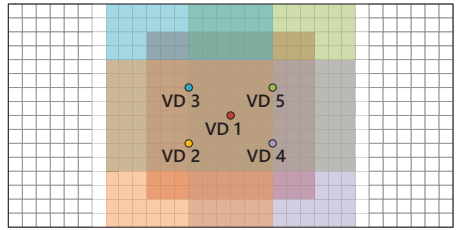}
\end{minipage}
}
\caption{ 5 possible viewing directions.}
\label{vd}
}
\end{figure}

In this section, we evaluate the proposed solutions in the four cases: the case without user transcoding and with the absolute smoothness requirement, i.e., case-(w/o,a), the case without user transcoding and with the relative smoothness requirement, i.e., case-(w/o,r), the case with user transcoding and with the absolute smoothness requirement, i.e., case-(w,a), and the case with user transcoding and with the relative smoothness requirement, i.e., case-(w,r).
In the simulation, we set $B=150$ MHz,\footnote{We consider a multi-carrier TDMA with 150 channels, each with bandwidth 1 MHz.} $T=50$ms, $P_{k}=2\times10^{-5}$ Watt, $U_{k}(r_{k})=r_{k},k\in\mathcal{K}$ and $n_{0}=Bk_{B}T_{0}$, where $k_{B}=1.38\times10^{-23}$ Joule/Kelvin is the Boltzmann constant and $T_{0}=300$ Kelvin is the temperature.
For ease of simulation, we consider two channel states for each user, i.e., a good channel state and a bad channel state,
and set $\mathcal{H}=\{d,2d\}$, $\mathrm{Pr}[H_{k}=d]=0.5$, and $\mathrm{Pr}[H_{k}=2d]=0.5$ for all $k\in\mathcal{K}$, where $d=10^{-6}$ reflects the path loss.
For ease of comparison, we set $\beta=1$ and refer to the performance metric in each case as energy for short.
We use \emph{Kvazaar} as the 360 VR video encoder and video sequence \emph{Reframe Iran} from YouTube as the video source. We set horizontal and vertical angular spans of each FoV as $100^{\circ}\times100^{\circ}$\cite{Ju:2017:UWV:3097895.3097899}. To avoid view switch delay in the presence of view changes, besides each requested FoV, we transmit an extra $10^{\circ}$ in every direction.
We set $M=18$, $N=36$ and $L=5$.
The encoding rates per tile and quantization parameters for the quality levels are shown in TABLE \ref{1}.
In addition, for ease of exposition, we consider $5$ possible viewing directions as shown in Fig. \ref{vd}, and assume that $K$ users randomly choose their viewing directions in an i.i.d. manner.
To capture the impact of the concentration of the viewing directions, assume $\mathcal{G}_{k},k\in\mathcal{K}$ follow a Zipf distribution, as in~\cite{8428401,8478317}.\footnote{Note that different 360 VR videos in general have different popularity distributions for viewing directions. Zipf distribution has been widely used to model content popularity in Internet and wireless networks. In particular, a larger Zipf exponent indicates a smaller tail of the popularity distribution, implying that Zipf exponent can reflect the concentration of requests for contents. In addition, note that the proposed solutions and their properties in this paper do not rely on a specific popularity distribution.}
In particular, the $v$-th popular viewing direction is chosen with probability $\frac{v^{-\gamma}}{\sum_{v\in\{1,...,5\}}v^{-\gamma}}$, where $v\in\{1,...,5\}$ and $\gamma$ is the Zipf exponent. When $\gamma$ is large (small), the concentration of the viewing directions of all users is high (low), implying that there are more (fewer) multicast opportunities.
We assume $r_{k},k\in\mathcal{K}$ follow the uniform distribution in $\{r_{lb},r_{lb}+1,...,r_{ub}\}$ with mean $\overline{r}=\frac{r_{lb}+r_{ub}}{2}$, where $r_{lb},r_{ub}\in\mathcal{L}$ and $r_{lb}<r_{ub}$.
We consider 200 random choices for $\mathcal{G}_{k},k\in\mathcal{K}$ and $r_{k},k\in\mathcal{K}$, and evaluate the average performance over these realizations.
\begin{table}[t]
\setlength{\abovecaptionskip}{0.1cm}
\centering
\caption{\label{1} Per tile encoding rates and quantization parameters for different quality levels.}
\setlength{\tabcolsep}{2mm}{
\begin{tabular}{|c|c|c|c|c|c|}
\hline
Quality level &1&2&3&4&5\\
\hline
Quantization parameter &42&35&28&21&14\\
\hline
Encoding rate ($\times10^{5}$bit/s) &6.66&16.18&24.29&32.01&40.23\\
\hline
\end{tabular}}
\end{table}

\begin{figure*}[t]
\normalsize{
\centering
\subfigure[ Energy versus $K$ at $\gamma=0$, $\Delta=1$, $r_{lb}=1$ and $r_{ub}=5$.]{
\begin{minipage}{6.5cm}
\centering
\includegraphics[width=6.5cm]{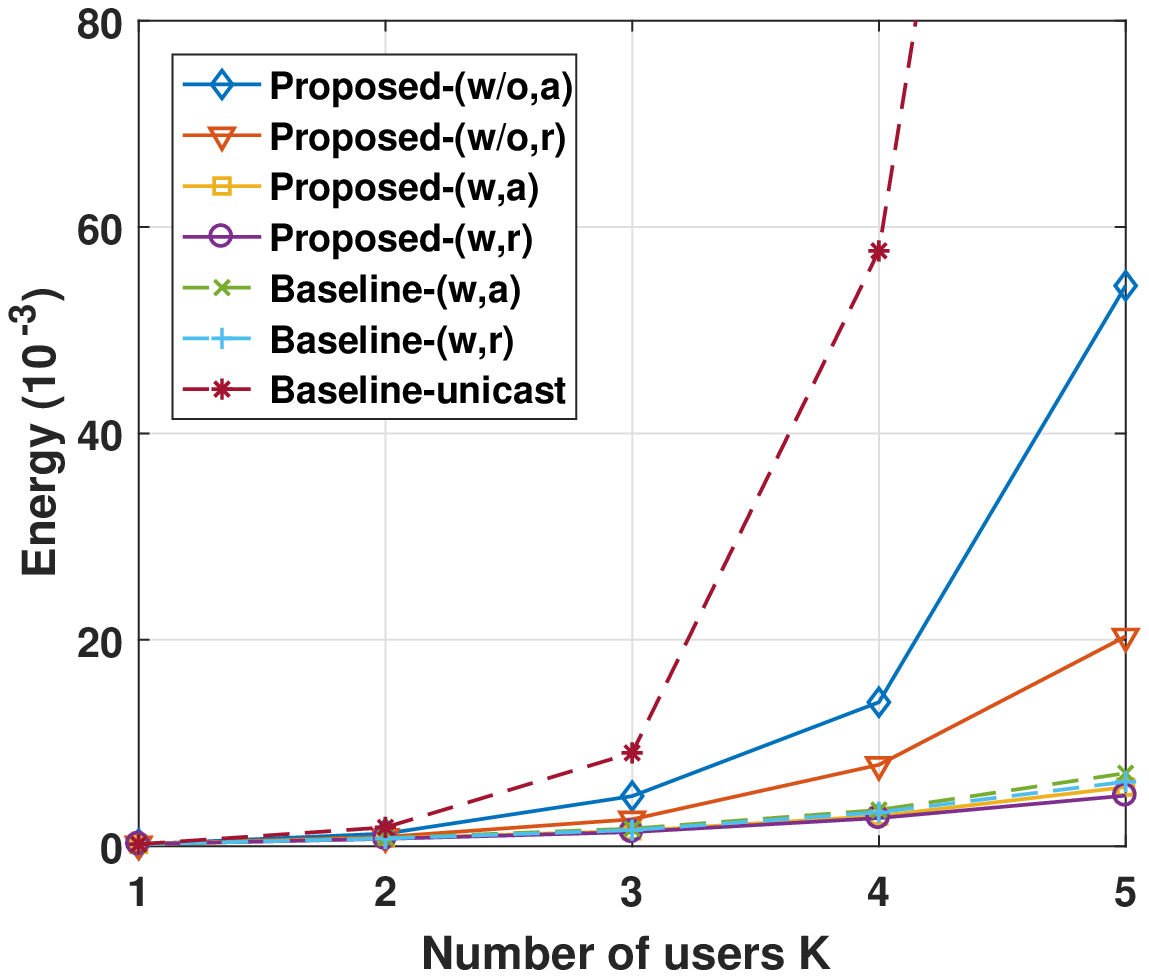}
\end{minipage}
}
\subfigure[ Energy versus $\overline{r}$ at $\gamma=0$, $K=3$, $\Delta=1$, and $r_{ub}=r_{lb}+2$. ]{ 
\begin{minipage}{6.5cm}
\centering
\includegraphics[width=6.5cm]{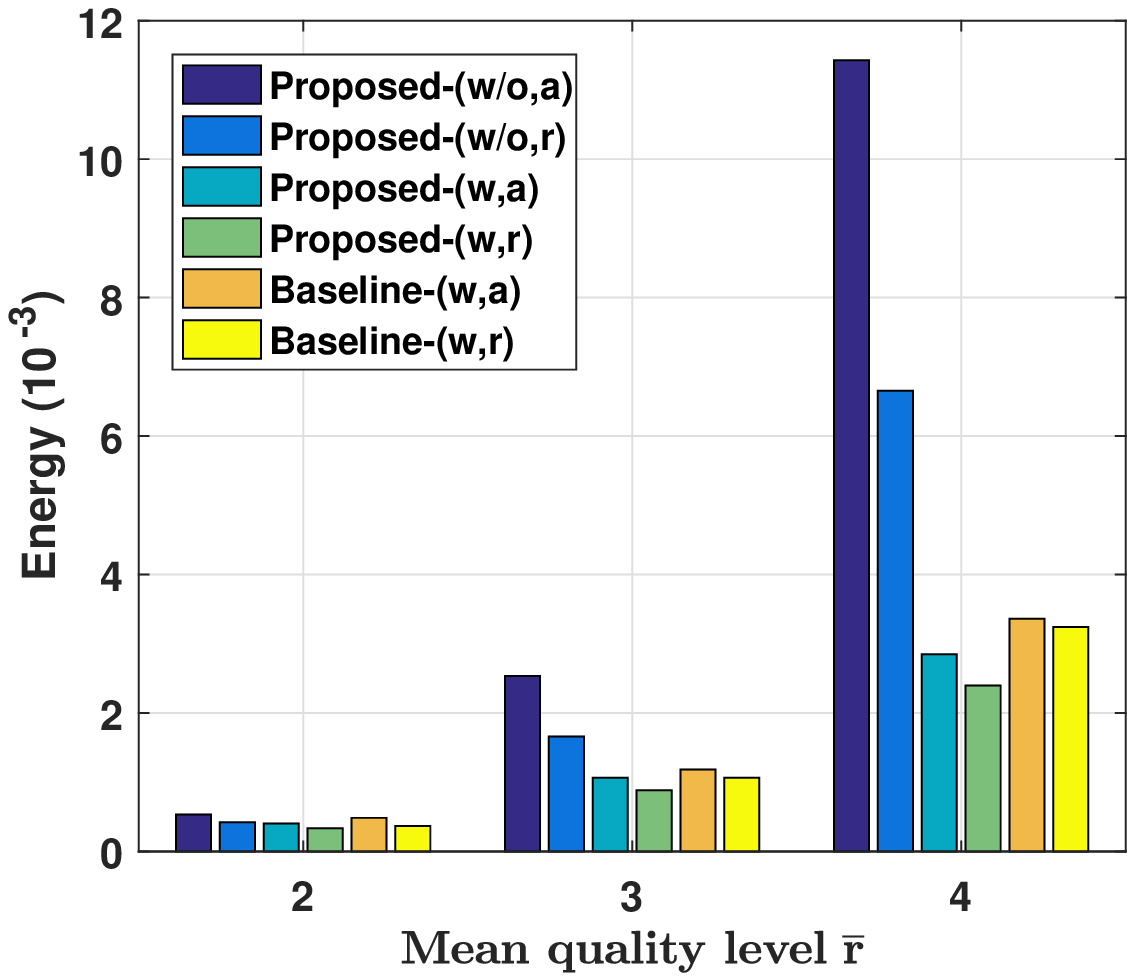}
\end{minipage}
}
\subfigure[ Energy versus $\gamma$ at $K=3$, $\Delta=1$, $r_{lb}=1$ and $r_{ub}=5$. ]{ 
\begin{minipage}{6.5cm}
\centering
\includegraphics[width=6.5cm]{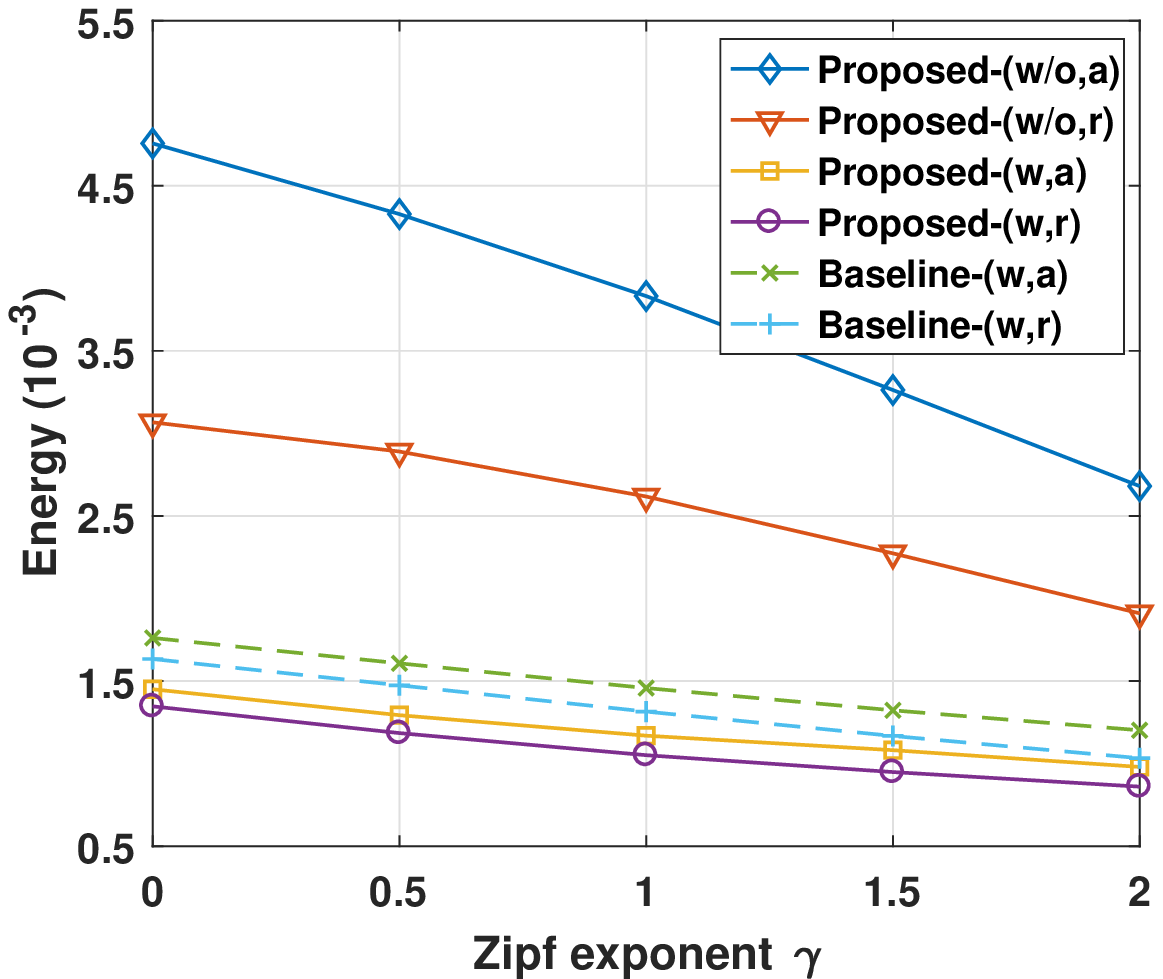}
\end{minipage}
}
\subfigure[ Energy versus $\Delta$ at $\gamma=0$, $K=3$, $r_{lb}=1$ and $r_{ub}=5$. ]{ 
\begin{minipage}{6.5cm}
\centering
\includegraphics[width=6.5cm]{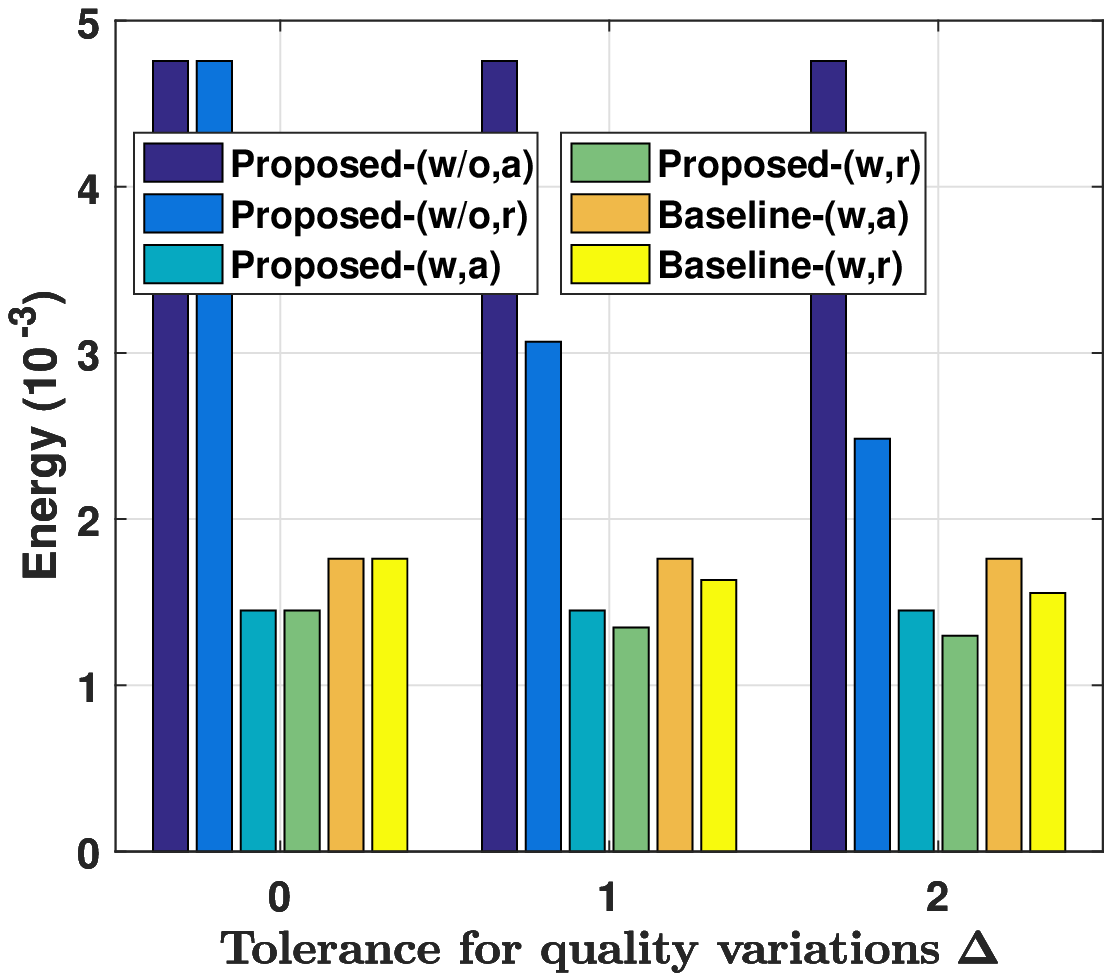}
\end{minipage}
}
\caption{ Energy comparison between the proposed solutions and baseline schemes. }
\label{sim1}
}
\end{figure*}

We consider three baseline schemes, namely Baseline-unicast, Baseline-(w,a) and Baseline-(w,r).
In Baseline-unicast, all users are served separately using unicast no matter whether $\mathcal{G}_{k},\:k\in\mathcal{K}$ are disjoint or not, and optimal power and time allocation is conducted by solving a problem similar to Problem~\ref{bP1}.
In Baseline-(w,a) and Baseline-(w,r), the $r_{\mathcal{S},\max}$-th representations of the tiles in $\mathcal{P}_{\mathcal{S}}$ are transmitted to all users in $\mathcal{S}$ using multicast, where $r_{\mathcal{S},\max}\triangleq \max_{k\in\mathcal{S}}r_{k}$.
In Baseline-(w,a), any user $k\in\mathcal{S}$ with $r_{k}=r_{\mathcal{S},\max}$ directly plays the received $r_{\mathcal{S},\max}$-th representations of the tiles in $P_{\mathcal{S}}$; any user $k\in\mathcal{S}$ with $r_{k}<r_{\mathcal{S},\max}$ converts the $r_{\mathcal{S},\max}$-th representations of the tiles in $P_{\mathcal{S}}$ to the $r_{k}$-th representations, and then plays them;
the corresponding optimal power and time allocation is obtained by solving Problem~\ref{P1} with $\mathbf{y}$ given by
\begin{align}
y_{\mathcal{S},k,l}=
\begin{cases}
1,& l= r_{\mathcal{S},\max}\\
0,& l\neq r_{\mathcal{S},\max}
\end{cases},\quad\mathcal{S}\in\mathcal{I},\ k\in\mathcal{S}.\label{basey}
\end{align}
In Baseline-(w,r), any user $k\in\mathcal{S}$ with $r_{\mathcal{S},\max}\leq r_{k}+\Delta$ directly plays the received $r_{\mathcal{S},\max}$-th representations of the tiles in $P_{\mathcal{S}}$; any user $k\in\mathcal{S}$ with $r_{\mathcal{S},\max}> r_{k}+\Delta$ converts the $r_{\mathcal{S},\max}$-th representations of the tiles in $P_{\mathcal{S}}$ to the $(r_{k}+\Delta)$-th representations, and then plays them;
the corresponding optimal power and time allocation is obtained by solving Problem~\ref{ssP1} with $\mathbf{y}$ given by \eqref{basey}.
Note that Baseline-unicast does not exploit any multicast opportunities and can be used in each case;
Baseline-(w,a) exploits natural multicast opportunities and transcoding-enabled multicast opportunities, and is applicable for case-(w,a);
Baseline-(w,r) exploits all three types of multicast opportunities, and can be utilized in case-(w,r).

Fig.~\ref{sim1} illustrates the energy versus the number of users $K$, the mean quality level $\overline{r}$, Zipf exponent $\gamma$ and the tolerance for quality variation $\Delta$.
From Fig.~\ref{sim1} (a), we can see that Baseline-unicast is much worse than the other schemes, revealing the importance of utilizing multicast opportunities in reducing energy consumption. To better compare the other competitive schemes, in the remaining figures, we no longer show the energy of Baseline-unicast which is much higher.
From Fig.~\ref{sim1} (a) and Fig.~\ref{sim1} (b), we can see that the energy of each scheme increases with $K$ and with $\overline{r}$, as the traffic load increases with $K$ and with $\overline{r}$.
From Fig.~\ref{sim1} (c), we can see that the energy of each scheme that utilizes multicast opportunities decreases with $\gamma$, due to the increment of multicast opportunities with $\gamma$.
From Fig.~\ref{sim1} (d), we can see that the energy of each scheme that utilizes relative smoothness-enabled multicast opportunities decreases with $\Delta$, due to the increment of relative smoothness-enabled multicast opportunities with $\Delta$.
Furthermore, from Fig.~\ref{sim1}, we can make the following observations. Proposed-(w/o,r) outperforms Proposed-(w/o,a), Proposed-(w,r) outperforms Proposed-(w,a), indicating that the relative smoothness requirement corresponds to more multicast opportunities than the absolute smoothness requirement.
Proposed-(w,a) outperforms Proposed-(w/o,a) and Proposed-(w,r) outperforms Proposed-(w/o,r), demonstrating that user transcoding can create multicast opportunities.
Proposed-(w,a) outperforms Baseline-(w,a) and Proposed-(w,r) outperforms Baseline-(w,r), showing the importance of optimally exploiting transcoding-enabled multicast opportunities in reducing energy consumption.

\section{Conclusion}\label{section_6}
In this paper, we investigated optimal wireless streaming of a multi-quality tiled 360 VR video to multiple users in wireless networks, by effectively utilizing characteristics of multi-quality tiled 360 VR videos and computation resources at the users’ side. In particular, we considered two requirements for quality variation in one FoV, i.e., the absolute smoothness requirement and the relative smoothness requirement, and two video playback modes, i.e., the direct-playback mode and transcode-playback mode. Besides natural multicast opportunities, we introduced two new types of multicast opportunities, i.e., relative smoothness-enabled multicast opportunities, and transcoding-enabled multicast opportunities, and established a novel mathematical model that reflects their impacts on the average transmission energy and transcoding energy. Then, we optimized the transmission resource allocation, playback quality level selection and transmission quality level selection to minimize the energy consumption in the four cases with different requirements for quality variation and video playback modes, by maximally exploiting potential multicast opportunities. By comparing the optimal values in the four cases, we proved that the energy consumption reduces when more multicast opportunities can be utilized. Finally, numerical results demonstrated the importance of effective exploitation of the three types of multicast opportunities. To the best of our knowledge, this is the first work that successfully utilizes relative smoothness-enabled multicast opportunities (with an effective guarantee for quality variation in each FoV) and transcoding-enabled multicast opportunities for optimal transmission of a multi-quality tiled 360 VR video to multiple users in wireless networks.

\begin{appendices}
\section*{ Appendix A: Proof of Theorem~\ref{lemma1}}
First, we relax the coupling constraints in \eqref{bxte1} and obtain the partial Lagrange function
$L(\mathbf{t},\mathbf{e},\boldsymbol{\lambda})\nonumber\triangleq \sum_{\mathbf{h}\in\mathcal{H}^{K}}\sum_{\mathcal{S}\in\mathcal{I}}\sum_{l\in\mathcal{L}}q_{\mathbf{H}}(\mathbf{h})e_{\mathbf{h},\mathcal{S},l}
+\sum_{\mathcal{S}\in\mathcal{I}}\sum_{k\in\mathcal{S}}\lambda_{\mathcal{S},k}\big(|\mathcal{P}_{\mathcal{S}}|D_{r_{k}}- \frac{B}{T}\mathbb{E}\big[t_{\mathbf{H},\mathcal{S},r_{k}}\log_{2}\big(1+\frac{e_{\mathbf{H},\mathcal{S},r_{k}}H_{k}}{t_{\mathbf{H},\mathcal{S},r_{k}}n_{0}}\big)\big]\big)
=\sum_{\mathbf{h}\in\mathcal{H}^{K}}q_{\mathbf{H}}(\mathbf{h})L_{\mathbf{h}}(\mathbf{t_{h}},\mathbf{e_{h}},\boldsymbol{\lambda})$,
where $\lambda_{\mathcal{S},k},\mathcal{S}\in\mathcal{I},k\in\mathcal{S}$ denote the Lagrange multipliers with respect to the constraints in $\eqref{bxte1}$ and
$L_{\mathbf{h}}(\mathbf{t_{h}},\mathbf{e_{h}},\boldsymbol{\lambda})\triangleq\sum_{\mathcal{S}\in\mathcal{I}}\sum_{l\in\mathcal{L}}e_{\mathbf{h},\mathcal{S},l}-\sum_{\mathcal{S}\in\mathcal{I}}\sum_{k\in\mathcal{S}}\lambda_{\mathcal{S},k}\big(\frac{B}{T}t_{\mathbf{h},\mathcal{S},r_{k}}\log_{2}\big(1+\frac{e_{\mathbf{h},\mathcal{S},r_{k}}h_{k}}{t_{\mathbf{h},\mathcal{S},r_{k}}n_{0}}\big)-|\mathcal{P}_{\mathcal{S}}|D_{r_{k}}\big).$
Next, we obtain the corresponding partial dual function of Problem~\ref{bP2}:
\begin{align}
D(\boldsymbol{\lambda})\triangleq
\mathop{\min}_{\mathbf{t},\mathbf{e}}& \quad
L(\mathbf{t},\mathbf{e},\boldsymbol{\lambda})\nonumber\\
\text{s.t.}&\quad \eqref{t1}, \eqref{t2}, \eqref{pi}, \eqref{ei1}.\nonumber
\end{align}
As the objective function and constraints are separable, this problem can be equivalently decomposed into the optimization in~\eqref{P33}, one for each $\mathbf{h}\in\mathcal{H}^{K}$.
As the duality gap for Problem~\ref{bP2} is zero, we can show Theorem~\ref{lemma1}.

\section*{ Appendix B: Proof of Theorem~\ref{lemma2}}
We prove Theorem~\ref{lemma2} by contradiction.
Suppose $(\mathbf{{x}}^{(\text{w,r})\star},\mathbf{{y}}^{(\text{w,r})\star})$ does not satisfy \eqref{lem4}.
That is, there exists $\mathcal{S}'\in\mathcal{I},k'\in\mathcal{K}$ such that
\begin{align}
x_{\mathcal{S}',k'}^{(\text{w,r})\star}\neq\min\left\{{r}^{(\text{w,r})}_{k'}+\Delta, \sum_{l\in\mathcal{L}}l{y}^{(\text{w,r})\star}_{\mathcal{S}',k',l}\right\}.\label{pro001}
\end{align}
As $(\mathbf{{x}}^{(\text{w,r})\star},\mathbf{{y}}^{(\text{w,r})\star})$ satisfies \eqref{0xr3} and \eqref{xiyk}, we know
\vspace{-2mm}
\begin{align}
x_{\mathcal{S},k}^{(\text{w,r})\star}\leq\min\left\{{r}^{(\text{w,r})}_{k}+\Delta, \sum_{l\in\mathcal{L}}l{y}^{(\text{w,r})\star}_{\mathcal{S},k,l}\right\},\quad \mathcal{S}\in\mathcal{I},\ k\in \mathcal{S}.\label{pro20}
\end{align}
By \eqref{pro001} and \eqref{pro20}, we know for some $\mathcal{S}'\in\mathcal{I},k'\in\mathcal{K}$,
\vspace{-2mm}
\begin{align}
x_{\mathcal{S}',k'}^{(\text{w,r})\star}<\min\left\{{r}^{(\text{w,r})}_{k'}+\Delta, \sum_{l\in\mathcal{L}}l{y}^{(\text{w,r})\star}_{\mathcal{S}',k',l}\right\}.\label{pro201}
\end{align}
Construct $\mathbf{{x}}^{\ddag}\triangleq({x}^{\ddag}_{\mathcal{S},k})_{\mathcal{S}\in\mathcal{I},k\in\mathcal{S}}$, where
\begin{align}
{x}^{\ddag}_{\mathcal{S},k}=\min\left\{{r}^{(\text{w,r})}_{k}+\Delta, \sum_{l\in\mathcal{L}}l{y}^{(\text{w,r})\star}_{\mathcal{S},k,l}\right\},\quad\mathcal{S}\in\mathcal{I},\ k\in\mathcal{S}.\label{pro200}
\end{align}
By \eqref{0xr3} and \eqref{xiyk}, we know
\begin{align}
\sum\nolimits_{l\in\mathcal{L}}l{y}^{(\text{w,r})\star}_{\mathcal{S},k,l}\geq r_{k},\quad \mathcal{S}\in\mathcal{I},\ k\in \mathcal{S}.\label{pro21}
\end{align}
By \eqref{pro200}, \eqref{pro21}, and the fact that $\mathbf{{y}}^{(\text{w,r})\star}$ satisfies \eqref{x0} and \eqref{x1}, we know that $(\mathbf{{x}}^{\ddag},\mathbf{{y}}^{(\text{w,r})\star})$ satisfies \eqref{0x0}, \eqref{0xr3} and \eqref{xiyk}.
Thus, $(\mathbf{{x}}^{\ddag},\mathbf{{y}}^{(\text{w,r})\star},\mathbf{{t}}^{(\text{w,r})\star},\mathbf{{p}}^{(\text{w,r})\star})$ is a feasible solution of Problem~\ref{0ssP1}.
In addition, by \eqref{pro201} and \eqref{pro200}, $x_{\mathcal{S}',k'}^{(\text{w,r})\star}<{x}^{\ddag}_{\mathcal{S}',k'}$ for some $\mathcal{S}'\in\mathcal{I},k'\in\mathcal{K}$. Thus, the objective value of Problem~\ref{0ssP1} at $(\mathbf{{x}}^{\ddag},\mathbf{{y}}^{(\text{w,r})\star},\mathbf{{t}}^{(\text{w,r})\star},\mathbf{{p}}^{(\text{w,r})\star})$ is smaller than the optimal value.
Therefore, by contradiction, we complete the proof of Theorem~\ref{lemma2}.

\section*{ Appendix C: Proof of Theorem~\ref{compare0}}
First, we show that ${E}^{(\text{w/o,r})\star}\leq E^{(\text{w/o,a})\star}$.
Construct $\mathbf{y}^{\ddag}\triangleq(y^{\ddag}_{\mathcal{S},k,l})_{\mathcal{S}\in\mathcal{I},k\in\mathcal{S},l\in\mathcal{L}}$, where
\begin{align}
y^{\ddag}_{\mathcal{S},k,l}=
\begin{cases}
1,& l= r_{k}\\
0,& l\neq r_{k}
\end{cases},\quad\mathcal{S}\in\mathcal{I},k\in\mathcal{S}.\label{pro1}
\end{align}
It is clear that $\mathbf{y}^{\ddag}$ satisfies \eqref{x0}, \eqref{x1} and \eqref{xiyk1}.
By \eqref{pro1} and the fact that $(\mathbf{t}^{(\text{w/o,a})\star},\mathbf{p}^{(\text{w/o,a})\star})$ satisfies \eqref{bxtp1},
we can show that $(\mathbf{y}^{\ddag},\mathbf{t}^{(\text{w/o,a})\star},\mathbf{p}^{(\text{w/o,a})\star})$ satisfies \eqref{xtp1}.
As an optimal solution of Problem~\ref{bP1}, $(\mathbf{t}^{(\text{w/o,a})\star},\mathbf{p}^{(\text{w/o,a})\star})$ satisfies \eqref{t1}, \eqref{t2} and \eqref{pi}.
Thus, $(\mathbf{y}^{\ddag},\mathbf{t}^{(\text{w/o,a})\star},\mathbf{p}^{(\text{w/o,a})\star})$ is a feasible solution of Problem~\ref{sP1}, with objective value $E^{(\text{w/o,a})\star}$. Thus, we have ${E}^{(\text{w/o,r})\star}\leq E^{(\text{w/o,a})\star}$.

Then, we show that ${E}^{(\text{w,r})\star}\leq{E}^{(\text{w,a})\star}$.
Construct $\mathbf{{x}}^{\ddag}\triangleq({x}^{\ddag}_{\mathcal{S},k})_{\mathcal{S}\in\mathcal{I},k\in\mathcal{S}}$, where
\begin{align}
{x}^{\ddag}_{\mathcal{S},k}=r_{k},\quad \mathcal{S}\in\mathcal{I},\ k\in \mathcal{S}.\label{pro3}
\end{align}
It is clear that $\mathbf{{x}}^{\ddag}$ satisfies \eqref{0x0} and \eqref{0xr3}.
By \eqref{pro3} and the fact that $\mathbf{{y}}^{(\text{w/o,r})\star}$ satisfies \eqref{xiyk01}, we can show that $(\mathbf{{x}}^{\ddag},\mathbf{{y}}^{(\text{w,a})\star})$ satisfies \eqref{xiyk}.
As an optimal solution of Problem~\ref{P1}, $(\mathbf{{y}}^{(\text{w,a})\star},\mathbf{{t}}^{(\text{w,a})\star},\mathbf{{p}}^{(\text{w,a})\star})$ satisfies \eqref{x0}, \eqref{x1}, \eqref{t1}, \eqref{t2}, \eqref{pi} and \eqref{xtp1}.
Thus, $(\mathbf{{x}}^{\ddag},\mathbf{{y}}^{(\text{w,a})\star},\mathbf{{t}}^{(\text{w,a})\star},\mathbf{{p}}^{(\text{w,a})\star})$ is a feasible solution of Problem~\ref{0ssP1}, with objective value ${E}^{(\text{w,a})\star}$. Thus, we have ${E}^{(\text{w,r})\star}\leq{E}^{(\text{w,a})\star}$.

Next, we show that ${E}^{(\text{w,a})\star}\leq E^{(\text{w/o,a})\star}$.
Construct $\mathbf{y}^{\ddag}\triangleq(y^{\ddag}_{\mathcal{S},k,l})_{\mathcal{S}\in\mathcal{I},k\in\mathcal{S},l\in\mathcal{L}}$ with $y^{\ddag}_{\mathcal{S},k,l}$ given by \eqref{pro1}.
It is clear that $\mathbf{y}^{\ddag}$ satisfies \eqref{x0}, \eqref{x1} and \eqref{xiyk}.
By \eqref{pro1} and the fact that $(\mathbf{t}^{(\text{w/o,a})\star},\mathbf{p}^{(\text{w/o,a})\star})$ satisfies \eqref{bxtp1},
we can show that $(\mathbf{y}^{\ddag},\mathbf{t}^{(\text{w/o,a})\star},\mathbf{p}^{(\text{w/o,a})\star})$ satisfies \eqref{xtp1}.
As an optimal solution of Problem~\ref{bP1}, $(\mathbf{t}^{(\text{w/o,a})\star},\mathbf{p}^{(\text{w/o,a})\star})$ satisfies \eqref{t1}, \eqref{t2} and \eqref{pi}.
Thus, $(\mathbf{y}^{\ddag},\mathbf{t}^{(\text{w/o,a})\star},\mathbf{p}^{(\text{w/o,a})\star})$ is a feasible solution of Problem~\ref{P1}, with objective value
$E^{(\text{w/o,a})\star}$. Thus, we have ${E}^{(\text{w,a})\star}\leq E^{(\text{w/o,a})\star}$.

Finally, we show that ${E}^{(\text{w,r})\star}\leq{E}^{(\text{w/o,r})\star}$.
Construct $\mathbf{\bar{x}^{\ddag}}\triangleq(\bar{x}^{\ddag}_{\mathcal{S},k})_{\mathcal{S}\in\mathcal{I},k\in\mathcal{S}}$, where
\begin{align}
\bar{x}^{\ddag}_{\mathcal{S},k}=\sum\nolimits_{l\in\mathcal{L}}l{y}^{(\text{w/o,a})\star}_{\mathcal{S},k,l},\quad \mathcal{S}\in\mathcal{I},\ k\in \mathcal{S}.\label{pro2}
\end{align}
It is clear that $\mathbf{\bar{x}}^{\ddag}$ satisfies \eqref{0x0} and \eqref{xiyk}.
By \eqref{pro2} and the fact that $\mathbf{{y}}^{(\text{w/o,r})\star}$ satisfies \eqref{xiyk1}, we can show that $\mathbf{\bar{x}}^{\ddag}$ satisfies \eqref{0xr3}.
As an optimal solution of Problem~\ref{sP1}, $(\mathbf{{y}}^{(\text{w/o,r})\star},$ $\mathbf{{t}}^{(\text{w/o,r})\star},\mathbf{{p}}^{(\text{w/o,r})\star})$ satisfies \eqref{x0}, \eqref{x1}, \eqref{t1}, \eqref{t2}, \eqref{pi} and \eqref{xtp1}.
Thus, $(\mathbf{\bar{x}}^{\ddag},\mathbf{{y}}^{(\text{w/o,r})\star},\mathbf{{t}}^{(\text{w/o,r})\star},\mathbf{{p}}^{(\text{w/o,r})\star})$ is a feasible solution of Problem~\ref{0ssP1}, with objective value ${E}^{(\text{w/o,r})\star}$.
Thus, we have ${E}^{(\text{w,r})\star}\leq{E}^{(\text{w/o,r})\star}$.

Therefore, we complete the proof of Theorem~\ref{compare0}.
\end{appendices}


\end{document}